\documentclass[12pt]{article}
\pdfoutput=1
\usepackage{epsf,amsfonts,amssymb,epsfig,amsmath,mathtools,graphics,slashed}
\usepackage{hyperref,hep,graphicx,subfig}
\usepackage{rotating}
\usepackage{xcolor}
\usepackage{verbatim}
\allowdisplaybreaks

\newcommand{\nn}{\notag \\}

\makeatletter

\@addtoreset{equation}{section}
\makeatother

\begin{document}

\begin{titlepage}

\vfill

\begin{flushright}
DCPT-21/11
\end{flushright}

\vfill

\begin{center}
   \baselineskip=16pt
   {\Large\bf Dissipation in Holographic Superfluids}
  \vskip 1.5cm
  \vskip 1.5cm
      Aristomenis Donos$^1$, Polydoros Kailidis$^1$ and Christiana Pantelidou$^2$\\
   \vskip .6cm
      \begin{small}
      \textit{$^1$ Centre for Particle Theory and Department of Mathematical Sciences,\\ Durham University,
       Durham, DH1 3LE, U.K.}\\
        \textit{$^2$ School of Mathematics, Trinity College Dublin, Dublin 2, Ireland}
        \end{small}\\   
         
\end{center}

\vfill

\begin{center}
\textbf{Abstract}
\end{center}
\begin{quote}
We study dissipation in holographic superfluids at finite temperature and zero chemical potential. The zero overlap with the heat current allows us to isolate the physics of the conserved current corresponding to the broken global $U(1)$. By using analytic techniques we write constitutive relations including the first non-trivial dissipative terms. The corresponding transport coefficients are determined in terms of thermodynamic quantities and the black hole horizon data. By analysing their behaviour close to the phase transition we show explicitly the breakdown of the hydrodynamic expansion. Finally, we study the pseudo-Goldstone mode that emerges upon introducing a perturbative symmetry breaking source and we determine its resonant frequency and decay rate.
\end{quote}

\vfill

\end{titlepage}

\setcounter{equation}{0}

\section{Introduction}
Holography provides a powerful framework to study large classes of strongly coupled systems at finite temperature. One of its most useful applications concerns the study of phase transitions and broken phases of thermodynamic systems. Moreover, the duality allows us to study large-$N$ systems at all energy scales. This includes the hydrodynamic limit of finite temperature systems which is expected to be universal.

In this paper we will be interested in the long wavelength excitations of the Goldstone mode that emerges in superfluid phases of matter which spontaneously break a global $U(1)$ symmetry. One of our aims is to include the first non-trivial dissipative effects which will result to a finite decay rate for the expected ``second sound". The second aim of this paper is to introduce perturbative sources which explicitly break the global symmetry and write down the effective theory of the resulting pseudo-Goldstone mode. 

One important aspect of the phases we will consider is that they will be at zero chemical potential and electric charge. This will allow us to isolate the long wavelength dynamics of the condensate avoiding its mixing with the other hydrodynamic degrees of freedom of the system.

Holography has been extensively used to construct and study superfluid phases after their original discovery \cite{Gubser:2008px,Hartnoll:2008kx,Hartnoll:2008vx}. From the bulk point of view, the boundary theory global $U(1)$ is mapped to a gauged $U(1)$ symmetry acting on a complex scalar field.  The vector field that gauges the $U(1)$ in the bulk is the gravitational dual of the conserved field theory Noether current operator. Below a critical temperature $T_c$ the charged scalar develops a perturbative instability yielding a new branch of broken phase black holes.

The effective theory of the Goldstone mode in the limit of long wavelengths is a topic which has been explored before. The starting point is the non-dissipative fluids which were first described in the framework of the Landau-Tisza theory \cite{PhysRev.60.356,PhysRev.72.838}. Dissipative effects at finite temperature were subsequently studied in the literature starting from \cite{KHALATNIKOV198270,ISRAEL198179} and more recently in \cite{Bhattacharya:2011eea}, in the context of a systematic classification\footnote{See also \cite{Davison:2016hno} for more recent developments in the framework of superfluids.}.

Due to the absence of chemical potential and the presence of rotational and time reversal symmetry, the only information we will need to extract from our holographic theories is two transport coefficients. We will be able to express these in terms of horizon data related to finite temperature static configurations. This, is in addition to the charge and current susceptibilities which are fixed at the level of thermodynamics and don't capture dissipative dynamics. A particularly useful tool in extracting these coefficients will be the symplectic current for gravity \cite{Crnkovic:1986ex}. At a practical level, this can be seen as the generalisation of the Wronskian for ordinary differential equations. However, it is powerful enough to be applicable in situations where the spacetime is inhomogeneous. Given the explicit expressions for our transport coefficients, we will be able to study our system near the phase transition. As we will see, the speed of second sound will go to zero as we approach the transition and the mode will become diffusive. However, its diffusion constant is not continuously connected to that of the incoherent current pole which exists in the normal phase.

Apart from the purely spontaneous case, we will also consider a scenario in which the global $U(1)$ is an almost exact symmetry\footnote{So far, the interplay of weak explicit and spontaneous breaking of symmetries has been mainly considered in the context of translations in e.g. \cite{Andrade:2015iyf,Amoretti:2018tzw, Donos:2019hpp}.  For the case of global symmetries in the bulk see \cite{Donos:2019txg,Amoretti:2018tzw}.}. To implement the explicit breaking of the $U(1)$ in a controlled manner, we will introduce a perturbative static source for our complex operator that condenses at low temperatures. As a result, the corresponding current becomes only partially conserved and the propagating pseudo-Goldstone mode acquires a mass as well as a finite gap. Our analysis is based on a modified Ward identity for the now almost conserved electric current and is therefore applicable beyond the details of our specific model.

In section \ref{sec:setup} we discuss the class of holographic models we will employ in order to realise our scenario along with some relevant aspects of their thermodynamics. In section \ref{sec:LinearResponse} we construct the theory of hydrodynamics which captures the dynamics of our Goldstone mode along with the technical aspects of our computation. We conclude the section with the computation of the Green's functions and the Kubo formulae for the two transport coefficients that enter our description. In section \ref{sec:explicit_breaking} we introduce perturbative static sources which break the global symmetry in a controlled manner as well as spacetime dependent ones for the complex scalar operator. We use the sourced Ward identity for the currents to extract the retarded Green's functions of the system. From the poles we extract the dispersion relations of second sound pseudo-Goldstone mode. Section \ref{sec:numerics} is devoted to numerical checks of some of the analytic results we obtain in the previous sections.

\section{Setup}\label{sec:setup}
In this section we will introduce our holographic superfluids.  Our main results on the hydrodynamics of the superfluidity Goldstone mode are independent of the specific scenario in which we will realise the neutral superfluid phase transition.  For concreteness, we will consider a holographic CFT which is deformed by introducing a source $\phi_{(s)} $ for the neutral relevant operator $\mathcal{O}_{\phi}$ with bulk dual $\phi$. Below a critical temperature $T_c$, which is set by the scale $\phi_{(s)} $, a boundary operator $\mathcal{O}_{\psi}$ which is charged under a global $U(1)$ symmetry condenses. From the bulk point of view, this condensation leads to black holes with non-zero profile for the dual bulk scalar $\psi$.

To realise this setup in holography, we consider a gravitational theory in the bulk consisting of a metric,  a neutral scalar $\phi$ and a complex scalar $\psi$ which is charged under a local $U(1)$ symmetry and a Maxwell field $A_{\mu}$ gauging the symmetry.  The local symmetry in the bulk corresponds to a global $U(1)$ symmetry on the boundary theory and the gauge field $A_\mu$ is dual to the corresponding Noether current operator $\hat{\mathcal{J}}^{\mu}$.

The system is described by the bulk action
\begin{align}\label{eq:baction}
S_{\mathrm{bulk}}=\int d^{4}x\,\sqrt{-g}\,\left(R-V(\phi,|\psi|^{2})-\frac{1}{2}\partial_{\mu}\phi\,\partial^{\mu}\phi-(D_{\mu}\psi)(D^{\mu}\psi)^{\ast}-\frac{1}{4}\tau(\phi,|\psi|^{2})\,F^{\mu\nu}F_{\mu\nu} \right)\,,
\end{align}
with the covariant derivative $D_{\mu}\psi=\nabla_{\mu}\psi+iqA_{\mu}\,\psi$ and the field strength $F=dA$. We will be mostly interested in configurations which are dual to superfluids and therefore spontaneously break the $U(1)$ symmetry. These backgrounds will have complex scalar with non-trivial modulus and for such configurations it will be beneficial to perform the field redefinition $\psi=\rho\,e^{i\theta}$. This brings the action to the equivalent form
\begin{align}\label{eq:baction2}
S=\int d^{4}x\,\sqrt{-g}\,\left(R-V(\phi,\rho^{2})-\frac{1}{2}(\partial\phi)^{2}-(\partial\rho)^{2}-\rho^{2}\,(\partial\theta+q A)^{2}-\frac{1}{4}\tau(\phi,\rho^{2})\,F^{\mu\nu}F_{\mu\nu} \right)\,.
\end{align}
The corresponding equations of motion are
\begin{align}\label{eq:eom}
R_{\mu\nu}-\frac{1}{2}g_{\mu\nu}V-\frac{\tau}{2}\left( F_{\mu\rho}F_{\nu}{}^{\rho}-\frac{1}{4}g_{\mu\nu}\,F^{2}\right) \qquad\qquad\qquad\qquad\qquad\qquad&\nn
-\frac{1}{2}\partial_{\mu}\phi\,\partial_{\nu}\phi-\partial_{\mu}\rho\,\partial_{\nu}\rho-\rho^{2}(\partial_{\mu}\theta+qA_{\mu})\,(\partial_{\nu}\theta+qA_{\nu})=&0\nn
\nabla_{\mu}\nabla^{\mu}\phi-\partial_{\phi}V-\frac{1}{4}\partial_\phi\tau\,F^{2}=&0\nn
\nabla_{\mu}\nabla^{\mu}\rho-\partial_{\rho^{2}}V\,\rho-\frac{1}{4}\partial_{\rho^{2}}\tau\,\rho\,F^{2}-\rho\,(\partial\theta+qA)^{2}=&0\nn
\nabla_{\mu}\left(\rho^{2}\left(\nabla^{\mu}\theta+q\,A^{\mu} \right)\right)=&0\nn
\nabla_{\mu}(\tau\,F^{\mu\nu})-2q\,\rho^{2}\,(\nabla^{\nu}\theta+q\,A^{\nu})=&0\,.
\end{align}

In order to implement our holographic scenario, we will assume that for small $\phi$ and $\psi$ the potential and the gauge coupling admit the expansions
\begin{align}\label{eq:V_Tau_Exp}
V&\approx -6+\frac{1}{2}m_\phi^2\,\phi^2+m_\psi^2\,|\psi|^2+\cdots\notag\\
\tau&\approx 1+\cdots\,.
\end{align}
Under these conditions, our theory admits the geometry of $AdS_4$ as solution with metric
\begin{align}
ds^2=r^2\left(- dt^2+dx^2+dy^2\right)+\frac{dr^2}{r^2}\,
\end{align}
and trivial scalars and gauge field. Given the expansion \eqref{eq:V_Tau_Exp}, the bulk scalars $\phi$ and $\psi$ correspond to operators $\mathcal{O}_\phi$ and $\mathcal{O}_\psi$ whose dimensions $\Delta_\phi$ and $\Delta_\psi$ are fixed by the mass terms according to $\Delta_\phi(\Delta_\phi-3)=m_\phi^2$ and $\Delta_\psi(\Delta_\psi-3)=m_\psi^2$. 

Without loss of generality, a suitable ansatz which captures all the necessary ingredients is
\begin{align}\label{eq:background}
ds^{2}&=-U(r)\,dt^{2}+\frac{dr^{2}}{U(r)}+e^{2g(r)}\,\left(dx^{2}+dy^{2} \right)\,,\nn
\phi&=\phi(r),\qquad \rho=\rho(r),\qquad A=0\,.
\end{align}
This ansatz leads to a non-linear system of ODEs for the radial functions that appear in it. Notice that the above choice of coordinates does not fully fix the radial coordinate $r$ which we are still free to shift by an arbitrary constant. We will choose to fix this freedom by requiring that the horizon of Hawking temperature $T$ is located at $r=0$. This allows us to write the near horizon expansion
\begin{align}\label{eq:bhor_exp}
&U(r)\approx\,4\pi T\,r+\mathcal{O}(r^{2}),\qquad g(r)\approx g^{(0)}+\mathcal{O}(r)\,,\nn
&\phi(r)\approx \phi^{(0)}+\mathcal{O}(r),\qquad \rho(r)\approx \rho^{(0)}+\mathcal{O}(r)\,.
\end{align}
Ultimately,  the backgrounds we wish to consider correspond to thermal states of the theory deformed by a source of the relevant operator $\mathcal{O}_\phi$ and a spontaneous VEV for the charged operator $\mathcal{O}_\psi$. Given these considerations, the appropriate near conformal boundary expansions at $r\to\infty$ are
\begin{align}\label{eq:UV_expansion}
&U(r)\approx (r+R)^{2}+\cdots+g_{(v)}\,(r+R)^{-1}+\cdots\,,\notag\\
&g(r)\approx \ln(r+R)+\dots\,,\nn
&\phi(r)\approx \phi_{(s)}\,(r+R)^{\Delta_{\phi}-3}+\cdots +\phi_{(v)}\,(r+R)^{-\Delta_{\phi}}+\cdots\,,\nn
&\rho(r)\approx \rho_{(s)}\,(r+R)^{\Delta_{\psi}-3}+\cdots\rho_{(v)}\,(r+R)^{-\Delta_{\psi}}+\cdots\,,
\end{align}
where we have the leading behaviour as well as the various constants of integration for our system. Given that we are looking for phases where $\mathcal{O}_{\psi}$ takes a VEV spontaneously, we will either set the sources $\rho_{(s)}$ equal to zero or, as we will do in section \ref{sec:explicit_breaking}, keep them as perturbative deformations. The VEVs of the scalar operators then are fixed by $\phi_{(v)}$ and $\rho_{(v)}$\footnote{In general, holographic renormalisation can involve the sources $\rho_{(s)}$ as part of the VEV of our scalar operators. However, we consider phases in which our operator takes a VEV spontaneously and we will naturally be working in the limit $\rho_{(s)}\ll \rho_{(v)}^{(3-\Delta_\psi)/\Delta_\psi}$. In this situation, the VEV will be approximately fixed by $\rho_{(v)}$.}. The constant shift $R$ is an artefact of the way we chose to fix our radial coordinate by placing the horizon at $r=0$.

At this point, it is worth understanding the asymptotics of the phase field $\theta$. In the absence of (or for perturbatively small) source $\rho_{(s)}$ a perturbation of $\delta\theta$ admits the asymptotic expansion
\begin{align}\label{eq:thetav}
\delta\theta(r)\approx r^{2\Delta_\psi-3}\,\delta\theta_{(s)}+\cdots+\delta\theta_{(v)}+\cdots\,.
\end{align}
In the complete absence of a source for $\psi$, the relevant bit for us is the constant term in $r$,  which parametrises the phase of the order parameter. Indeed, for the VEV of the condensed operator we will have that $\langle\mathcal{O}_\psi \rangle\propto \rho_{(v)}\,e^{i\,\theta_{(v)}}$.  From the boundary point of view, we can simply change the global phase without any cost in energy. The fact that we can freely rotate the VEV $\langle\mathcal{O}_\psi \rangle$ without changing the energy of our system reflects the existence of a Goldstone mode in the boundary theory.  This suggests that the dynamics of the Goldstone mode is captured by the VEV of the operator
\begin{align}
\mathcal{O}_{Y}=\frac{1}{2 i\,\left|\langle\mathcal{O}_\psi \rangle_b\right|}\,(\langle\mathcal{O}_{\bar{\psi}} \rangle_b\,\mathcal{O}_{\psi}-\langle\mathcal{O}_\psi \rangle_b\,\mathcal{O}_{\bar{\psi}})\,,
\label{eq:OY_def}
\end{align}
where $\langle\mathcal{O}_\psi \rangle$ is the thermal state VEV.  In the absence of a source for $\psi$, for the fluctuations we can write
\begin{align}\label{eq:vevOY}
\delta \langle\mathcal{O}_Y \rangle=\langle\mathcal{O}_\psi \rangle\,\delta\theta_{(v)}\,.
\end{align}

In section \ref{sec:explicit_breaking} of this paper will be interested in hydrodynamic perturbations in which the source $\delta\theta_{(s)}$ will be much smaller than the scale of $\delta\theta_{(v)}$. Given this information, in this language the perturbative source for the scalar will be $\delta s_{\psi}=i(\rho_{(v)}\,\delta \theta_{(s)}+\delta\rho_{(s)}\,\delta\theta_{(v)})$ while the VEV will be given by $\langle\mathcal{O}_\psi \rangle\propto (2\Delta_\psi-3)\,\rho_{(v)}\,(1+i\delta\theta_{(v)})$ after a perturbation. Therefore the source of the operator $\mathcal{O}_{Y}$ will be simply $s_Y=(\rho_{(v)}\,\delta \theta_{(s)}+\delta\rho_{(s)}\,\delta\theta_{(v)})$ while for its VEV we can write equation \eqref{eq:vevOY}.

A central point of interest to our paper is the fluctuations of the bulk gauge field $A$ around the background black holes of equation \eqref{eq:background}. From the equations of motion \eqref{eq:eom}, we see that fluctuations of the gauge field $A_{\mu}$ and phase field $\theta$ are captured by the last two equations. In fact, after defining the one-form field $B_{\mu}=\partial_{\mu}\theta+q\,A_{\mu}$, we can simply consider the equation of motion of a massive vector field in the bulk
\begin{align}\label{eq:beom}
\nabla_{\mu}(\tau\,W^{\mu\nu})-2q^{2}\rho^{2}\,B^{\nu}=0\,,
\end{align}
where we have defined the two form $W=dB$. The penultimate equation of \eqref{eq:eom} now reads
\begin{align}
\nabla_{\mu}\left( \rho^2\,B^\mu\right)=0\,,
\end{align}
and is simply a consequence of taking the divergence of equation \eqref{eq:beom}. For this reason we will only need to consider equation \eqref{eq:beom}. Close to the conformal boundary,  in the absence of background sources for complex scalar, the 1-form field $B_\nu$ admits the expansion
\begin{align}\label{eq:Basympt}
B_{\alpha}=\partial_\alpha\theta_{(s)}\,r^{2\Delta_\psi-3}+\cdots+v_{\alpha}+\cdots+\frac{q\,j_{\alpha}}{r}+\cdots\,,
\end{align}
where $v_{\alpha}=\partial_{\alpha}\theta_{(v)}+q\,\mu_{\alpha}$ is a gauge invariant combination of the superfluid velocity $\partial_{\alpha}\theta_{(v)}$ and the source $\mu_{\alpha}$ for the $U(1)$ current. Moreover, as we will explain later the constants of integration $j_\alpha$ satisfy the Ward identity for the currents which are given by equation \eqref{eq:ftcurrent}. Note that in the thermal states we are interested in we have $\mu_{\alpha}=0$; we will only consider non-zero $\mu_{\alpha}$ when computing the thermodynamic susceptibilities. 

\subsection{Thermodynamics}\label{sec:thermo}
An important ingredient in studying the thermodynamics and the linear response of our system is the free energy density $w$. In order to compute it, we first need to regularise our bulk action \eqref{eq:baction} by introducing appropriate couterterms which render the total action finite and the boundary value problem well defined. These form a boundary action which is defined on a hypersurface $\partial M$ of constant radial coordinate $r$ near the conformal boundary.  An appropriate choice of boundary action includes the terms
\begin{align}\label{eq:bdy_action}
S_{bdr}=&-\int_{\partial M}d^{3}x\,\sqrt{-\gamma}\,\left(-2K + 4 +R_{bdr}\right)\notag\\
&\quad -\frac{1}{2}\int_{\partial M}d^{3}x\,\sqrt{-\gamma}\,[(3-\Delta_{\phi})\phi^2-\frac{1}{2\Delta_{\phi}-5}\,\partial_{a}\phi\partial^{a}\phi]\notag\\
&\quad -\int_{\partial M}d^{3}x\,\sqrt{-\gamma}\,[(3-\Delta_{\psi})|\psi|^2-\frac{1}{2\Delta_{\psi}-5}\,D_{a}\psi D^{a}\psi^{\ast}]+\cdots\,,
\end{align}
with $\gamma_{\alpha\beta}$ the induced metric on $\partial M$, $R_{bdr}$ the associated Ricci scalar and $K$ the extrinsic curvature scalar of the constant $r$ hypersurface. In fact, for the case with $\Delta_{\phi}=\Delta_{\psi}=2$ on which we focus in our numerics section \ref{sec:numerics}, these are all the terms needed.

In order to compute the free energy we need to evaluate the regularised action $S_{tot}=S_{bulk}+S_{bdr}$ on the Euclidean version of our backgrounds \eqref{eq:background} with $t=-i\,\tau$. Since we are dealing with an infinite field theory system, this is an infinite quantity and we should be discussing densities instead. After dropping the integration over the boundary spatial non-compact coordinates, this yields the density $I_{tot}$ which is related to the free energy density $w_{FE}$ via $w_{FE}=T I_{tot}$ with $T$ the Hawking temperature of our black brane. Notice that, in the absence of sources, the terms which involve the complex scalar $\psi$ in \eqref{eq:bdy_action} will not contribute to the thermodynamics of our system. However, these are needed in order to extract the VEV of the complex operator $\mathcal{O}_{\psi}$.

In order to compute the on-shell value of the Euclidean bulk action $I_{bulk}$, it is useful to write the integrand as a total derivative. After exploiting the existence of the Killing vector $\partial_t$ and using a Komar type of argument, we can write
\begin{align}\label{eq:bonshell}
I_{bulk}=\frac{1}{T}\int_{0}^{\infty}dr\,\left(e^{2g}\,U^{\prime} \right)^{\prime}\,.
\end{align}
Being in the grand canonical ensemble and at zero background chemical potential, the free energy density of our system is
\begin{align}
w_{FE}=\epsilon- T\,s,
\end{align}
where $\epsilon$ is the energy density and $s$ is the Bekenstein-Hawking entropy. This can be expressed in terms of the horizon data \eqref{eq:bhor_exp} as
\begin{align}\label{eq:entropy_density}
s=4\pi\,e^{2\,g^{(0)}}\,.
\end{align}
In the case of e.g. $\Delta_{\phi}=\Delta_{\psi}=2$ the bulk on-shell action \eqref{eq:bonshell} combines with the boundary action \eqref{eq:bdy_action} to yield
\begin{align}
w_{FE}=-2g_{(v)}-\phi_{(s)}\,\phi_{(v)}-T\,s\,.
\end{align}

Even though we have not included a source $\mu_\alpha$ for the current, this is certainly an important part of thermodynanics for a superfluid. The response of the system to an external source includes a non-trivial expectation value for the conserved $U(1)$ current
\begin{align}\label{eq:ftcurrent}
J^\alpha=\lim_{r\to\infty}\sqrt{-g}\,\tau\,F^{\alpha r}=\lim_{r\to\infty}\sqrt{-g}\,\tau\,W^{\alpha r}\,.
\end{align}
For our purposes, we will need this information in the context of linear response.

In addition to the free energy $w$, we are also interested in the thermodynamic susceptibilities of our system. In order to compute them, we consider the static perturbations $\delta B^{(t)}$ and $\delta B^{(x)}$ with zero superfluid velocity. In terms of the asymptotics \eqref{eq:Basympt}, we have $\delta v_{\nu}=q\,\delta^{t}_{\nu}\,\delta \mu_{t}$ and $\delta v_{\nu}=q\,\delta^{x}_{\nu}\,\delta \mu_{x}$. Close to the conformal boundary these perturbations will admit the expansion
\begin{align}\label{eq:therm_pert_UV}
\delta B^{(t)}_{t}&=q\,\delta\mu_{t}+q\,\frac{\delta j_{t}}{r+R}+\cdots\,,\nn
\delta B^{(x)}_{x}&=q\,\delta\mu_{x}+q\,\frac{\delta j_{x}}{r+R}+\cdots\,.
\end{align}
In terms of the thermodynamic susceptibilities $\chi_{QQ}$ and $\chi_{JJ}$ we must have $\delta j_{t}=-\chi_{QQ}\,\delta\mu_{t}$\footnote{The charge density is $\delta\rho=-\delta j_{t}$.} and $\delta j_{x}=-\chi_{JJ}\,\delta\mu_{x}$. Close to the event horizon, regularity demands the expansion
\begin{align}\label{eq:therm_pert_IR}
\delta B^{(t)}_{t}&=q\, \delta\mu_{t}\, a_{t}^{(0)}\,r+\mathcal{O}(r^{2})\,,\nn
\delta B^{(x)}_{x}&=q\, \delta\mu_{x}\, a_{x}^{(0)}+\mathcal{O}(r)\,.
\end{align}
The quantities $a_{t}^{(0)}$ and $a_{x}^{(0)}$ are certainly part of the thermodynamics of a superfluid. Later, we will see that together with the horizon data of \eqref{eq:nh_exp} they will determine the transport coefficients of our superfluid when we express the current in terms of the superfluid velocity $\partial_\alpha\theta_{(v)}$.

Later in the paper, we will also want to understand our hydrodynamic expansion close to the phase transition. For the case where the transition is second order, close to the critical temperature $T_c$ we must have $\rho\propto \sqrt{T_c-T}$ corresponding to a VEV which scales like $\langle\mathcal{O}_\psi\rangle\propto \sqrt{T_c-T}$.  This will let us see the important fact that the Goldstone mode does not connect continuously to the diffusive mode of the incoherent current which exists above $T_c$. More interestingly, we will see that the source for this discontinuity is that our hydrodynamic expansion breaks down close to the critical temperature making the radius of convergence infinitesimally small.

By integrating the bulk equations of motion \eqref{eq:beom} we obtain the relations
\begin{align}
\delta \rho&=e^{2g^{(0)}}\tau^{(0)}\delta\mu_t\,a_t^{(0)}+2q\,\int_{0}^{\infty}dr\,\rho^2\frac{e^{2g}}{U}\delta B_{t}^{(t)}\,,\notag\\
\delta j_x&=-2q\,\int_{0}^{\infty}dr\,\rho^2\,\delta B_x^{(x)}\,.
\end{align}
As we can see, the right hand side of the above equations still involves the perturbation of the one-form field. Close to the phase transitions these equations yield
\begin{align}\label{eq:susc_approx}
\chi_{QQ}&=e^{2g^{(0)}}\tau^{(0)}\,a_t^{(0)}+\mathcal{O}\left(T_c-T\right)\,,\notag\\
\chi_{JJ}&=2q^2\,\int_0^\infty dr\,\rho^2+\mathcal{O}\left((T_c-T)^2\right)\,.
\end{align}
This is indeed the behaviour we would have expected close to a superfluid phase transition with the charge susceptibility $\chi_{QQ}$ remaining finite and the current susceptibility $\chi_{JJ}$ approaching zero.

\section{Linear Response}\label{sec:LinearResponse}

In this section, we will study long wavelength fluctuations of the $U(1)$ current of our system. The thermal state above the critical temperature is simply an electrically neutral hot plasma corresponding to a CFT that has been deformed by a relevant operator while preserving Poincare invariance. In the absence of electric charge, the electric current fluctuations are dominated by a diffusive mode in the hydrodynamic limit \cite{Kovtun:2003wp, Saremi:2007dn, Iqbal:2008by}. The existence of this mode in a neutral plasma is due to the existence of an incoherent current in strongly coupled CFTs \cite{Damle:1997rxu, Hartnoll:2007ih}.

Below the critical temperature the hydrodynamics of the electric current will be dominated by the Goldstone mode associated to the spontaneous breaking of the global $U(1)$ of the boundary theory. In section \ref{sec:sympl} we will introduce the symplectic current in the bulk which will be the main technical tool we will use for our computations. In section \ref{sec:sec_sound} we will construct the hydrodynamic perturbations which couple to the Goldstone mode of our boundary theory. This will allow us to write constitutive relations for the expectation value of the current operator in terms of the phase of our order parameter. Using these and current conservation we will extract the dispersion relation of the second sound mode in terms of thermodynamic susceptibilities and two transport coefficients which are expressed in terms of horizon quantities. Finally, in section \ref{sec:greens} we will include sources for the current in our analysis. We will give expressions for the retarded Green's functions of the system and Kubo formulae for the transport coefficients of our superfluid.

\subsection{Symplectic Current}\label{sec:sympl}

A powerful tool we will use is the symplectic current for the bulk theory. This will allow us to deal with the one-form field fluctuations in an elegant way. Its benefit lies in the fact that it is a conserved current in the bulk and will allow us to relate boundary quantities to horizon ones. As we will see, this logic will allow us to express the boundary electric current in terms of derivatives of the phase of our order parameter, the superfluid velocity, and the sources.

In order to introduce, it we imagine that we have two perturbations $\delta B^{<1>}_\mu$ and $\delta B^{<2>}_\mu$ which solve our bulk equation \eqref{eq:beom}. We can then construct the vector density
\begin{align}\label{eq:sympl_current}
P^{\mu}=\sqrt{-g}\,\tau\,\left(\delta B^{<1>}_{\nu}\delta (W^{<2>})^{\nu\mu}-\delta B^{<2>}_{\nu}\delta (W^{<1>})^{\nu\mu} \right)\,,
\end{align}
which can be shown to be divergence free,
\begin{align}\label{eq:divP}
\partial_{\mu}P^{\mu}=0\,.
\end{align}

This current is not a consequence of a conservation law as it is based on two independent solutions. It merely comes from the fact that the equations of motion of our classical theory in the bulk are derived from a local classical action. In the hydrodynamic limit, the dominant term in \eqref{eq:divP} will be the one involving the radial derivative which will let us relate the symplectic current on the black brane horizon to the symplectic current on the conformal boundary boundary.

This is especially helpful when one of the two solutions we are using in the construction in the construction of \eqref{eq:sympl_current} is considered to be known. For our purposes, the role of the known solution will be played by the static perturbations $\delta B^{(t)}_\mu$ and $\delta B^{(x)}_\mu$ we discussed in section \ref{sec:thermo}. At a philosophical level, we will need to consider as known all the static black holes which can be used to describe the thermodynamics of our system. That would include the black holes corresponding to finite chemical potential as well as persistent supercurrents.  The reason is that, as we will see, the susceptibilities $\chi_{QQ}$ and $\chi_{JJ}$ will play an important role even though we will be studying transport in thermal states of zero chemical potential and external vector field.  For the purpose of extracting those, the knowledge of the static perturbations $\delta B^{(t)}_\mu$ and $\delta B^{(x)}_\mu$ is sufficient.

\subsection{Second sound}\label{sec:sec_sound}

In this section we would like to construct the bulk perturbation corresponding to the Goldstone mode of the boundary theory: the second sound. As we discussed in section \ref{sec:thermo}, the Goldstone mode involves the phase of our order parameter and therefore the associated conserved current due to the global symmetry on the boundary. In the absence of electric charge, the current decouples from the stress tensor of the theory within linear response. This will allow us to clearly isolate the dynamics of the Goldstone mode from the rest of the hydrodynamics modes of our system.

From the bulk point of view we will only need to examine perturbations of the one-form field $B_\mu$. The spacetime translational symmetries allow us to study Fourier modes. Thus, in order to study perturbations in the hydrodynamic limit, we will consider long wavelength excitations of the form
\begin{align}\label{eq:qnm_ansatz}
\delta B_{\mu}(t,x;r)=e^{-i\omega (t+S(r))+i\varepsilon k\,x}\,\delta b_{\mu}(r)\,dx^{\mu}\,,
\end{align}
for some small number $\varepsilon$.  The function $S(r)$ is chosen so that it drops faster that $\mathcal{O}(1/r^3)$ close to the conformal boundary and it therefore doesn't interfere with holographic renormalisation. However, close to the horizon, it is chosen so that it approaches $S(r)\to \frac{1}{4\pi T}\ln r$ and the combination $t+S(r)$ is regular and ingoing. Note that we picked the momentum $k$ to point in the direction $x$ without loss of generality given that the background is isotropic.

An important ingredient in extracting the second sound is the absence of boundary sources. In general, the asymptotic expansion close to the conformal boundary is
\begin{align}
\delta b_{\alpha}=\delta \hat{v}_{\alpha}+q\,\frac{\delta \hat{j}_{\alpha}}{r}+\cdots\,.
\end{align}
The leading terms are a gauge invariant combination of the external one-form source and the superfluid velocity. Absence of sources is equivalent to demanding
\begin{align}\label{eq:gbcs}
\delta \hat{v}_{t}=-i\omega\,\delta \hat{c},\qquad \delta \hat{v}_{x}=i\varepsilon k\,\delta \hat{c}\,,
\end{align}
so that $v_{\alpha}=\partial_{\alpha}\theta_{(v)}$ in the notation of equation \eqref{eq:Basympt}.
Moreover, charge conservation implies that
\begin{align}\label{eq:chargecont}
\partial_{\alpha}\delta j^{\alpha}=0\Rightarrow \omega \delta \hat{j}_{t}+ \varepsilon k \delta\hat{j}_{x}=0\,.
\end{align}

At this point it will be illuminating to count constants of integration. This is crucial in order to see that we are doing the right thing in terms of boundary conditions since we are after specific quasi-normal modes.  The equation of motion \eqref{eq:beom} yields two second order equations for $\delta b_t$ and $\delta b_x$ while $\delta b_r$ can be solved algebraically. In order to find a unique solution we will therefore need to fix four constants.

Close to the conformal boundary we have the constant $\delta \hat{c}$, while $\delta\hat{j}_t$ and $\delta\hat{j}_t$ give one more constant since they have to satisfy the constraint \eqref{eq:chargecont}. Close to the horizon, in-falling boundary conditions fix the expansion
\begin{align}
\delta b_\alpha&=\delta b_\alpha^{(0)}+r\,\delta b_\alpha^{(1)}+\cdots\,,\notag\\
\delta b_r&=\frac{1}{4\pi Tr}\delta b_t^{(0)}+\delta b_r^{(1)}+\cdots\,.
\end{align}
We therefore have an extra two constants of integration coming from the near horizon expansion, giving an overall total of 4 constants. However, we are only solving a linear system of homogeneous equations. Since we are not introducing any sources, we can use the scaling symmetry of the problem to set any of the above constants of integration to one leaving with only three. The fourth constant is precisely the frequency $\omega$ which will ultimately become a function of the wavenumber $k$ that we are free to choose. This procedure will fix the dispersion relation for $\omega(\epsilon k)$ for the quasi-normal modes.

For our purposes however, we are after a particular mode which is hydrodynamic and has $\omega=0$ for $\varepsilon=0$. This simply corresponds to the trivial solution for the one form field. This allows us to consider the hydrodynamic expansion
\begin{align}
\omega&=\varepsilon\,\omega_{[1]}+\varepsilon^{2}\,\omega_{[2]}+\cdots\,,\nn
\delta\hat{j}_{\mu}&=\varepsilon\,\delta\hat{j}_{\mu[1]}+\varepsilon^{2}\,\delta\hat{j}_{\mu[2]}+\cdots\,.
\end{align}
For the radial function in the ansatz \eqref{eq:qnm_ansatz} we will write the expansion
\begin{align}\label{eq:qnm_eps_exp}
\delta b_{t}(r)&=\varepsilon\,\delta \hat{B}^{(t)}_{t}(r)+\varepsilon^{2}\,\delta B^{(2)}_{t}(r)+\cdots\,,\nn
\delta b_{x}(r)&=\varepsilon\,\delta \hat{B}^{(x)}_{x}(r)+\varepsilon^{2}\,\delta B^{(2)}_{x}(r)+\cdots\,,\nn
\delta b_{r}(r)&=\varepsilon^{2}\,\delta  B^{(2)}_{r}(r)+\cdots\,,
\end{align}
where $\delta \hat{B}^{(t)}_{t}(r)$ and $\delta \hat{B}^{(x)}_{x}(r)$ are precisely the thermodynamic perturbations we discussed in equation \eqref{eq:Basympt} with
\begin{align}
\delta\mu_{t}=-\frac{i\omega_{[1]}}{q}\delta \hat{c},\qquad \delta\mu_{x}=\frac{i k}{q}\delta \hat{c}\,.
\end{align}
For the subleading terms we impose the near conformal boundary expansion
\begin{align}
\delta B_{t}^{(n)}&=-i\omega_{[n]}\delta \hat{c}+ q\frac{\delta\hat{j}_{t[n]}}{r}+\cdots\,,\nn
\delta B_{x}^{(n)}&= q\frac{\delta\hat{j}_{x[n]}}{r}+\cdots\,,
\end{align}
which is simply a reorganisation of the perturbation with boundary conditions as in equation \eqref{eq:gbcs}. Close to the horizon, we impose in-falling boundary conditions. This gives us the expansion,
\begin{align}\label{eq:nh_exp}
\delta B_{t}^{(n)}&= \delta B^{(n)(0)}_{t}+\delta B^{(n)(1)}_{t}\,r+\cdots\,,\nn
\delta B_{r}^{(n)}&= \frac{\delta B^{(n)(0)}_{t}}{4\pi Tr}+\delta B^{(n)(0)}_{r}+\cdots\,,\nn
\delta B_{x}^{(n)}&= \delta B^{(n)(0)}_{x}+\delta B^{(n)(1)}_{x}\,r+\cdots\,.
\end{align}

At leading order in the $\varepsilon$ expansion the components of the conserved current are
\begin{align}\label{eq:currents_leading}
q\,\delta\hat{j}_{t[1]}=i\omega_{[1]}\chi_{QQ}\,\delta \hat{c},\qquad q\,\delta\hat{j}_{x[1]}=-ik\chi_{JJ}\,\delta \hat{c}\,,
\end{align}
which follows from the definition of the charge and current susceptibilities. At that order, we see that equation \eqref{eq:chargecont} gives
\begin{align}\label{eq:disp_linear}
&\omega_{[1]}\delta \hat{j}_{t[1]}+k\delta \hat{j}_{x[1]}=0\Rightarrow \omega_{[1]}^{2}\,\chi_{QQ}-k^{2}\,\chi_{JJ}=0\nn
&\Rightarrow\omega_{[1]}=\pm\,c_{s}\,k\,,
\end{align}
with $c_{s}^{2}=\chi_{JJ}/\chi_{QQ}$ being the universal speed of second sound.

To specify $\omega_{[2]}$, we will need to fix the currents $\delta\hat{j}_{\alpha[2]}$.  In order to do this we will employ the conservation \eqref{eq:divP} of the symplectic current \eqref{eq:sympl_current}. As we explained in section \ref{sec:sympl}, the symplectic current is particularly useful when we compare the solution we are after to a solution we already know. For this reason we will consider it for two different choices of pairs of perturbations. For both cases, we will choose $\delta B^{<1>}$ to be the perturbation in \eqref{eq:qnm_ansatz}.

Our first choice for $\delta B^{<2>}$ is the static perturbation $\delta B^{(x)}$ of section \ref{sec:thermo}. This will allow us to specify the component $\delta\hat{j}_x$. Expanding the symplectic current in $\varepsilon$ we have
\begin{align}\label{eq:sympl_exp}
P^{\mu}=e^{-i\omega (t+S(r))+i\varepsilon k\,x}\,\left(\varepsilon^{2}\,p^{\mu}_{(2)}+\varepsilon^3\,p^{\mu}_{(3)}+\cdots \right)\,.
\end{align}
One would expect that the leading term would be $\mathcal{O}(\varepsilon)$ since our perturbation $\delta B^{<1>}$ starts at order $\varepsilon$ and $\delta B^{<2>}$ starts at zeroth order. However, the leading term in the $\varepsilon$-expansion of  $\delta B^{<1>}$ in equation \eqref{eq:qnm_eps_exp} is proportional to $\delta B^{<2>}$ and the would-be leading term in the symplectic current turns out to vanish.

More explicitly we have
\begin{align}
p^{t}_{(2)}&=\frac{i\tau}{U}\delta B^{(x)}_{x}\,\left(\omega_{[1]}\delta \hat{B}^{(x)}_{x}+k\delta \hat{B}^{(t)}_{t} \right)\,,\nn
p^{r}_{(2)}&=-U\tau\,\left(\delta B^{(2)}_{x}\,\partial_{r}\delta B^{(x)}_{x}+\delta B^{(x)}_{x}\,\left(i \omega_{[1]} S^{\prime}\, \delta \hat{B}^{(x)}_{x}-\partial_{r}\delta B^{(2)}_{x}\right) \right)\,,\nn
p^{x}_{(2)}&=U\tau\,\delta B_{r}^{(2)}\,\partial_{r}\delta B^{(x)}_{x} \,.
\end{align}
We now consider the divergence \eqref{eq:divP} and keep only terms up to $\mathcal{O}(\varepsilon^{2})$ giving
\begin{align}
\partial_{r}p_{(2)}^{r}=0\,.
\end{align}
Integrating this equation from the horizon up to the conformal boundary, we obtain the relation
\begin{align}\label{eq:current_2ndO}
q\,\delta \hat{j}_{x[2]}=-\omega_{[1]}k\,\sigma_{d}\,\delta \hat{c}\,,
\end{align}
where we have defined \footnote{For this transport coefficient, a similar result was obtained in \cite{Davison:2015taa}. Here we clarify the significance of the quantities that appear in this equation as data related to the geometries dual to the thermal state after the inclusion of the superfluid velocity in the ensemble variables.}
\begin{align}\label{eq:sigmad_def}
\sigma_d=\tau^{(0)}\,(a_x^{(0)})^2\,.
\end{align}

Choosing now $\delta B^{<2>}$ to be $\delta B^{(t)}$, we obtain
\begin{align}
p^{t}_{(2)}&=-e^{2g}\tau\,\delta B^{(2)}_{r}\,\partial_{r}\delta B^{(t)}_{t}\,,\nn
p^{r}_{(2)}&=e^{2g}\tau\,\left(\delta B^{(2)}_{t}\,\partial_{r}\delta B^{(t)}_{t}+\delta B^{(t)}_{t}\,\left(i \omega_{[1]} S^{\prime}\, \delta \hat{B}^{(t)}_{t}-\partial_{r}\delta B^{(2)}_{t}\right) \right)\,,\nn
p^{x}_{(2)}&=-\frac{i\tau}{U}\delta B^{(t)}_{t}\,\left( \omega_{[1]}\,\delta\hat{B}^{(x)}_{x}+k\,\delta\hat{B}^{(t)}_{t}\right) \,.
\end{align}
Keeping once again only terms up to order $\mathcal{O}(\varepsilon^{2})$ in equation \eqref{eq:divP} and integrating from the horizon up to the conformal boundary, we obtain
\begin{align}
q\,\delta\hat{j}_{t[2]}=i\omega_{[2]}\,\chi_{QQ}\,\delta \hat{c}+e^{2g^{(0)}}\tau^{(0)}\,\delta B_{t}^{(2)(0)}\,a_{t}^{(0)}\,.
\end{align}
We have used the near horizon expansion \eqref{eq:nh_exp} and we need to specify the constant of integration $\delta B_{t}^{(2)(0)}$. We can do this by considering the near horizon limit of the radial component of the equation of motion \eqref{eq:beom} along with the expansion \eqref{eq:nh_exp}. This fixes
\begin{align}
\delta B^{(2)(0)}_{t}=-\frac{\omega_{[1]}^{2}}{2q^{2}(\rho^{(0)})^{2}}a_{t}^{(0)}\tau^{(0)}\,\delta \hat{c}\,,
\end{align}
yielding
\begin{align}\label{eq:charge_2ndO}
q\,\delta\hat{j}_{t[2]}=i\omega_{[2]}\,\chi_{QQ}\,\delta \hat{c}-e^{2g^{(0)}}\,\frac{(\tau^{(0)}a_{t}^{(0)})^{2}}{2q^{2}(\rho^{(0)})^{2}}\omega_{[1]}^{2}\,\delta \hat{c}\,.
\end{align}

Moving on to next to leading order in equation \eqref{eq:chargecont} along with \eqref{eq:charge_2ndO} and \eqref{eq:current_2ndO}, we have
\begin{align}
\omega_{[2]}\,\delta\hat{j}_{t[1]}+\omega_{[1]}\,\delta\hat{j}_{t[2]}+k\,\delta\hat{j}_{x[2]}&=0\Rightarrow\nn
2i\,\chi_{QQ}\,\omega_{[2]}-e^{2g^{(0)}}\,\frac{(\tau^{(0)}a_{t}^{(0)})^{2}}{2q^{2}(\rho^{(0)})^{2}}\omega_{[1]}^{2}-k^{2}\sigma_{d}&=0\Rightarrow\nn
\omega_{[2]}=-i\frac{1}{2\chi_{QQ}}\left(e^{2g^{(0)}}\,\frac{(\tau^{(0)}a_{t}^{(0)})^{2}}{2q^{2}(\rho^{(0)})^{2}}\frac{\chi_{JJ}}{\chi_{QQ}}+\sigma_{d} \right)\,k^{2}&\,.
\end{align}

Equations \eqref{eq:currents_leading}, \eqref{eq:current_2ndO} and \eqref{eq:charge_2ndO} can be considered as the expansion of the constitutive relations for the conserved current. By introducing the Fourier transform $\delta c$ of $\delta \hat{c}$, we can write them in the form
\begin{align}\label{eq:jconst_rel}
q\,\delta j_{t}&=-\chi_{QQ}\,\partial_{t} \delta c+\Xi\,\partial_{t}^{2}\delta c\,,\nn
q\,\delta j_{x}&=-\chi_{JJ}\,\partial_{x}\delta c-\sigma_{d}\,\partial_{x}\partial_{t}\delta c\,,
\end{align}
where we have defined
\begin{align}\label{eq:xi_def}
\Xi=e^{2g^{(0)}}\,\frac{(\tau^{(0)}a_{t}^{(0)})^{2}}{2q^{2}(\rho^{(0)})^{2}}\,.
\end{align}
In addition to the constitutive relations for the currents, the VEV of our charged scalar operator is
\begin{align}\label{eq:scalar_const}
\langle\mathcal{O}_\psi\rangle=\langle\mathcal{O}_\psi\rangle_b\,(1+i\,\delta c)\,,
\end{align}
at leading order in the $\varepsilon$ expansion. Here, we have introduced the VEV of the scalar in the thermal state $\langle\mathcal{O}_\psi\rangle_b=(2\Delta-3)\,\rho_{(v)}$. 

Using this notation, the dispersion relation for our sound mode is
\begin{align}\label{eq:disp_rel}
\omega=\pm\,c_s\,k-\frac{i}{2\chi_{QQ}^2}\left(\chi_{JJ}\,\Xi+\chi_{QQ}\,\sigma_d \right)\,k^2\,.
\end{align}
Note that these constitutive relations are based on the two thermodynamic susceptibilities $\chi_{QQ}$ and $\chi_{JJ}$ as well as the two transport coefficients $\sigma_d$ and $\Xi$. In the next section we will introduce sources for the external vector field which will allow us to compute the retarded Green's functions of the current. As a result, we will manage to write Kubo formulae for our transport coefficients $\sigma_d$ and $\Xi$. 

Before moving on, we would like to examine the behaviour of our hydrodynamics close to the critical temperature $T_c$. As one can see from the constitutive relations \eqref{eq:jconst_rel_sources}, the coefficients that enter in our hydrodynamic expansion include the susceptibilities $\chi_{JJ}$ and $\chi_{QQ}$. The behaviour of these coefficients close to $T_c$ was given in equation \eqref{eq:susc_approx}. The first derivative with respect to the temperature of both of them exhibits the expected discontinuity close to the transition with the former approaching zero. From that, we can easily see that the speed of second sound fixing the linear part of the dispersion relation\eqref{eq:disp_linear} behaves like $c_s\approx (T_c-T)^{1/2}$ close to the phase transition. 

The dissipative part of the constitutive relations \eqref{eq:jconst_rel} is determined by the transport coefficients $\sigma_d$ and $\Xi$.  The holographic expressions that we obtained for these are given in equations \eqref{eq:sigmad_def} and \eqref{eq:xi_def} respectively.  The behaviour of $\sigma_d$ is fairly simple to understand as $\tau^{(0)}$ goes to a fixed value as we increase temperature and $a_x^{(0)}$ is simply unity at $T_c$.  On the other hand,  equation \eqref{eq:xi_def} is telling us that $\Xi$ blows up close to the transition since $\rho^{(0)}\propto (T_c-T)^{1/2}$ and everything else remains finite in our holographic expression. We interpret this fact as the breakdown of the hydrodynamic expansion making the radius of convergence smaller and smaller as we approach phase transition. One would have expected such a breakdown since the phase $\delta c$ becomes a not well defined field close to the transition. This is easy to see directly from the definition of $\mathcal{O}_Y$ itself in equation \eqref{eq:OY_def}. Note that there is another mode in the theory which becomes gapless close to the transition which is associated with the modulus of the complex VEV $\langle\mathcal{O}_\psi\rangle$ \cite{Bhaseen:2012gg}. However, this mode completely decouples from the system we are examining as it is captured by the bulk field $\rho$ in our parametrisation.

Despite this fact, the speed of sound and the attenuation in the dispersion relation \eqref{eq:disp_rel} remain finite since, according to equation \eqref{eq:susc_approx}, the current-current scalars behave like $\chi_{JJ}\approx (T_c-T)$. We can easily see that close to the transition the dispersion relation \eqref{eq:disp_linear} becomes
\begin{align}
\omega\approx -\frac{i}{2}\left(\frac{4\pi}{s}\,\int_0^{\infty}dr\,\frac{\rho^2}{(\rho^{(0)})^2} +\frac{\sigma_d}{\chi_{QQ}}\right)\,k^2\,,
\end{align}
which remains finite as we take the $T\to T_c$ limit. Note that this dispersion relation holds for $k\ll T_c-T$.

Above the phase transition where our system is in its normal phase, the $U(1)$ current hydrodynamic excitations are given by the constitutive relations,
\begin{align}
\delta j_t&=-\chi_{QQ}\,\delta\mu\,,\notag\\
\delta j_x&=-\sigma_d\,\partial_x\delta\mu\,,
\end{align}
with $\delta\mu$ the local chemical potential and $\sigma_d$ the incoherent conductivity. We therefore see that above the critical temperature we will have a single diffusive mode determined by the diffusion constant $D_{inc}=\chi_{QQ}^{-1}\,\sigma_d$. So, even though our sound modes tend to become diffusive close to the transition, the attenuation constant does not continuously connect to the diffusion constant of the normal phase incoherent mode at the critical temperature.

\subsection{Green's functions}\label{sec:greens}

In this section we will introduces sources for the conserved $U(1)$ current in our system which should be of order $\delta s_{\alpha}\approx \mathcal{O}(\varepsilon)$. The construction of the hydrodynamic perturbation in the bulk is identical to that of equation \eqref{eq:qnm_eps_exp}. However, this time the frequency is fixed by the external sources and our ultimate goal is to express the boundary current of the system in terms of the sources and the phase of the condensed operator.

The above suggests that we will simply need to replace the boundary conditions \eqref{eq:gbcs} by
\begin{align}\label{eq:gfbcs}
\delta \hat{v}_{t}=\delta\hat{s}_t-i\omega\,\delta \hat{c},\qquad \delta \hat{v}_{x}=\delta\hat{s}_x+i\varepsilon k\,\delta \hat{c}\,,
\end{align}
with $\delta\hat{s}_\alpha\propto\mathcal{O}(\varepsilon)$. Effectively, the whole analysis of section \ref{sec:sec_sound} goes through after replacing $-i\omega\,\delta\hat{c}\to \delta\hat{s}_t-i\omega\,\delta \hat{c}$ and $i\varepsilon k\,\delta \hat{c}\to \delta\hat{s}_x+i\varepsilon k\,\delta \hat{c}$. In terms of spacetime coordinates on the boundary, this is equivalent to replacing the partial derivatives of the phase by the gauge invariant combination
\begin{align}
\partial_\alpha\delta c\to \delta s_\alpha+\partial_\alpha\delta c\,.
\end{align}

One can of course check this explicitly but here we will just state the final result for the expressions of the components of the boundary current
\begin{align}\label{eq:jconst_rel_sources}
q\,\delta j_{t}&=-\chi_{QQ}\,\left(\delta s_t+\partial_{t} \delta c\right)+\Xi\,\partial_{t}\left(\delta s_t+\partial_{t} \delta c\right) \,,\nn
q\,\delta j_{x}&=-\chi_{JJ}\,\left( \delta s_x+\partial_{x}\delta c\right)-\sigma_{d}\,\partial_{t}\left( \delta s_x+\partial_{x}\delta c \right)\,.
\end{align}
After having obtained the constitutive relations \eqref{eq:jconst_rel_sources}, we can impose the continuity equation \eqref{eq:chargecont} to obtain an equation for the Goldstone mode. Doing this in momentum space, gives us an expression for $\delta\hat{c}$ in terms of the sources $\delta\hat{s}_\alpha$ which we can plug back in \eqref{eq:jconst_rel_sources}. From that linear relation between the VEVs for the current and the sources we read off the retarded Green's functions
\begin{align}\label{eq:Greens}
G_{J^t J^t}(\omega,\varepsilon k)&=(\varepsilon k)^2\,\mathcal{G}(\omega,\varepsilon k),\qquad G_{J^x J^x}(\omega,\varepsilon k)=\omega^2\,\mathcal{G}(\omega,\varepsilon k)\,,\notag\\
G_{J^t J^x}(\omega,\varepsilon k)&=G_{J^x J^t}(\omega,\varepsilon k)=\varepsilon k\,\omega\,\mathcal{G}(\omega,\varepsilon k)\,,
\end{align}
where we have defined
\begin{align}\label{eq:mathcalG}
\mathcal{G}(\omega,\varepsilon k)=\frac{\left(\chi_{QQ}+i\,\Xi\,\omega\right)\left( i\,\chi_{JJ}+\sigma_d\,\omega\right)}{\omega^2\left( -i\,\chi_{QQ}+\Xi\,\omega\right)+(\varepsilon k)^2\left(i\,\chi_{JJ}+\sigma_d\,\omega\,\right)}\,.
\end{align}
Notice that the last equation in \eqref{eq:Greens} is compatible with the Onsager relation $G_{J^t J^x}(\omega,\varepsilon k)=-G_{J^x J^t}(\omega,-\varepsilon k)$ given that $\mathcal{G}$ is even in $k$. Moreover, the positions of poles of the Green's functions \eqref{eq:Greens} are set by the roots of the denominator of the function \eqref{eq:mathcalG}. As expected, these are located precisely on the curves $\omega=\omega_\pm(\varepsilon k)$ set by the dispersion relations we found in section \ref{sec:sec_sound} for the superfluid sound mode.

The final result we would like to present in this section is the Kubo formulae for our transport coefficients $\sigma_d$ and $\Xi$. By using the Green's functions \eqref{eq:Greens} we can write
\begin{align}
\sigma_d=\lim_{\omega\to 0}\lim_{k\to 0} \frac{\mathrm{Im}\, G_{J^x J^x}}{\omega},\qquad \Xi=\lim_{k\to 0}\lim_{\omega\to 0} \frac{\mathrm{Im}\, G_{J^t J^t}}{\omega}\,.
\end{align}

\section{Scalar Sources and Pinning}\label{sec:explicit_breaking}

In this section we would like to explicitly deform our theory by a perturbatively small pinning parameter $\delta\rho_{(s)}$ and study the resulting pseudo-Goldstone mode. In other words, we would like to explicitly break the global $U(1)$ in a controlled fashion. In order to do this, we will have to deform the backgrounds we have considered so far by adding a small perturbative source $\rho_{(s)}=\delta\rho_{(s)}\propto\mathcal{O}(\varepsilon^2)$ in the near conformal boundary expansion of equation \eqref{eq:UV_expansion}.

Before applying our logic to the class of holographic theories we are considering, it is worth revisiting our expectations in field theory terms. Suppose that we couple our theory to an external gauge field $A_\alpha$ and that we also introduce a source $\lambda$ for our charged scalar operators $\mathcal{O}_\psi^\ast$. If the resulting theory is invariant under the infinitesimal gauge transformations
\begin{align}
\delta A_\alpha=\partial_\alpha\delta\Lambda,\qquad \delta \lambda=-iq\lambda\,\delta\Lambda\,,
\end{align}
then we can show that the corresponding Ward identity for the current gets modified to
\begin{align}\label{eq:current_non_cons}
\nabla_\alpha\langle J^\alpha\rangle=iq\,\left(\langle \mathcal{O}_\psi\rangle \lambda^\ast-\langle \mathcal{O}_\psi^\ast\rangle \lambda\right)\,.
\end{align}
This equation makes clear that the order parameter and the source need to align in the complex plane in equilibrium. In other words we can no longer freely rotate the order parameter in the complex plane after fixing a source without generating a current.

In holography, the small background perturbation we want to consider changes the interpretation of the angle $\theta_{(v)}$ that we introduced in equation \eqref{eq:thetav}. To see this, we need to consider the asymptotic expansion of the charged scalar close to the conformal boundary. To do this, we first need to better understand the asymptotics of the vector field $B_\mu$ to include sources $\theta_{(s)}$ for the Goldstone mode. In the absence of background sources for the charged scalar field, we have
\begin{align}\label{eq:B_exp_sources}
B_{\alpha}=\partial_\alpha\theta_{(s)}\,(r+R)^{2\Delta-3}+\cdots+v_{\alpha}+\cdots+q\,j_\alpha\,(r+R)^{-1}+\cdots\,,
\end{align}
where $\theta_{(s)}$, $v_\alpha$ and $j_\alpha$ are constants of integration, with the equations of motion implying the constraint
\begin{align}\label{eq:current_non_conservation}
\partial_\alpha j^\alpha=2q\,\rho_{(v)}^2\,(2\Delta-3)\,\theta_{(s)}+\mathcal{O}(\partial^2\theta_{(s)})\,,
\end{align}
where we  dropped terms of higher order in derivatives of $\theta_{(s)}$.  The reason for doing this is that we will be interested in satisfying this equation up to order $\mathcal{O}(\varepsilon^3)$ and such terms would be higher order provided that $\theta_{(s)}\propto\mathcal{O}(\varepsilon^2)$. We will show this later when we consider the proper boundary conditions, compatible with the absence of time dependent sources for the charged scalar operator $\mathcal{O}_\psi$.

After including the perturbative source terms, the angle $\theta$ and the background source term $\delta\rho_{(s)}$ enter the asymptotic expansion of the charged scalar perturbations according to,
\begin{align}
\delta\psi=i(\rho_{(v)}\,\theta_{(s)}+\delta\rho_{(s)}\theta_{(v)})\,(r+R)^{\Delta-3}+\cdots+i\rho_{(v)}\,\theta_{(v)}\,(r+R)^{-\Delta}+\cdots\,.
\end{align}
This shows that equation \eqref{eq:current_non_conservation} is nothing but the Ward identity \eqref{eq:current_non_cons} after identifying
\begin{align}
\langle J^\alpha\rangle&=j^\alpha+\mathcal{O}(\partial\theta_{(s)}),\notag\\
\lambda&=\delta\rho_{(s)}+i(\rho_{(v)}\,\theta_{(s)}+\delta\rho_{(s)}\theta_{(v)})+\mathcal{O}(\delta\rho_{(s)}\,\theta_{(s)}),\notag\\
\langle\mathcal{O}_\psi\rangle&=\langle\mathcal{O}_\psi\rangle_b\,(1+i\theta_{(v)})+\mathcal{O}(\delta\rho_{(s)}\,\theta_{(s)})\,.
\end{align}
From the above we see that in order to correctly identify the time dependent perturbative source $\delta s_\psi$ for the scalar field, we need to impose
\begin{align}
\theta_{(s)}=\frac{1}{\rho_{(v)}}\left(-\delta\rho_{(s)}\,\theta_{(v)}+\delta s_\psi\right)\,.
\end{align}
Note that in this notation, the time dependent source of the complex scalar introduces only a source $\delta s_Y=\delta s_\psi$ for the operator $\mathcal{O}_{Y}$ that we introduced in section \ref{sec:setup}. The above equation shows that the source term $\theta_{(s)}$ for the perturbation of $B_\mu$ and the background perturbation source $\delta\rho_{(s)}$ need to be of the same order in the $\varepsilon$ expansion. After this observation, the constraint equation \eqref{eq:current_non_conservation} becomes
\begin{align}\label{eq:current_non_conservationV2}
\partial_\alpha j^\alpha=-2\,q\,|\langle\mathcal{O}_\psi\rangle_b|\,\delta\rho_{(s)}\,\theta_{(v)}+2\,q\,|\langle\mathcal{O}_\psi\rangle_b|\,\delta s_Y\,.
\end{align}

From the point of view of hydrodynamics, the goal is to have the correct constitutive relations for the currents $j_\alpha$ up to second order in $\varepsilon$. That would allow us to satisfy equation \eqref{eq:current_non_conservationV2} up to third order in $\varepsilon$ provided that we take the source for the charged scalar to be of order $\mathcal{O}(\varepsilon^2)$. This is justified by the fact that the terms we dropped in equation \eqref{eq:current_non_conservation} are of order $\mathcal{O}(\varepsilon^4)$. This argument shows that the constitutive relations of equations \eqref{eq:jconst_rel_sources} and \eqref{eq:scalar_const} are sufficient for this task. Finally, one might worry that we should take in account the fact that the background quantities which enter equation \eqref{eq:beom} will bring their own $\varepsilon$ corrections to the perturbation of $B_\mu$. However, as we argued above, the correction of the bulk scalar due to the perturbative source $\delta\rho_{(s)}$ will be of order $\mathcal{O}(\varepsilon^2)$. That would induce corrections to our perturbation $\delta B_\mu$ at order $\mathcal{O}(\varepsilon^3)$ which is certainly beyond our scope.

In order to compute the retarded Green's functions of the system of our operators, we need to solve the Ward identity \eqref{eq:current_non_conservationV2} after identifying $\theta_{(v)}=\delta c$. By doing so we obtain the explicit expressions
\begin{align}\label{eq:greens_pinning}
G_{J^t J^t}(\omega,\varepsilon k)&=\frac{f\,(w+(\varepsilon\, k)^2\,g)}{h}\,,\quad G_{J^x J^x}(\omega,\varepsilon k)=\frac{(-w+\omega^2 f)\,g}{h}\notag\\
G_{J^x J^t}(\omega,\varepsilon k)&=G_{J^t J^x}(\omega,\varepsilon k)=\varepsilon\, k\,\omega\frac{f\,g}{h}\,,\notag\\
G_{J^t \mathcal{O}_Y}(\omega,\varepsilon k)&=-i\frac{q\omega\,|\langle\mathcal{O}_\psi\rangle_b|\,f}{h}\,,\quad G_{J^x \mathcal{O}_Y}(\omega,\varepsilon k)=-i\frac{q k\,|\langle\mathcal{O}_\psi\rangle_b|\,g}{h}\notag\\
G_{\mathcal{O}_Y J^t}(\omega,\varepsilon k)&=i\frac{q\omega\,|\langle\mathcal{O}_\psi\rangle_b|\,f}{h}\,,\quad G_{\mathcal{O}_Y J^x}(\omega,\varepsilon k)=i\frac{q k\,|\langle\mathcal{O}_\psi\rangle_b|\,g}{h}\notag\\
G_{\mathcal{O}_Y \mathcal{O}_Y}(\omega,\varepsilon k)&=\frac{q^2\,|\langle\mathcal{O}_\psi\rangle_b|^2}{h}
\end{align}
where we have defined
\begin{align}
f&=\chi_{QQ}+i\,\Xi\,\omega,\qquad g=\chi_{JJ}-i\,\sigma_d\,\omega\,,\notag\\
h&=w-\omega^2\,f+\varepsilon^2\,k^2\,g,\qquad w=2q^2\,|\langle\mathcal{O}_\psi\rangle_b|\,\delta\rho_{(s)}\,.
\end{align}
The poles of the Green's functions in equation \eqref{eq:greens_pinning} reveal that the second sound mode has acquired a resonance frequency as well as a gap. More specifically, we find the dispersion relation
\begin{align}\label{eq:gap}
\omega=\pm\sqrt{\frac{w+k^2\,\chi_{JJ}}{\chi_{QQ}}}-\frac{i}{2\chi_{QQ}^2}\left( w\,\Xi+k^2\,(\Xi\,\chi_{JJ}+\sigma_d\,\chi_{QQ})\right)\,,
\end{align}
from which we can read off the resonance frequency $\omega_r=\sqrt{w/\chi_{QQ}}$ and the gap\footnote{Notice that in the language of \cite{Donos:2019txg}, we have $\omega_{gap}\sim \chi_{YY}^{-1}$ with $\chi_{YY}$ the susceptibility of the operator $\mathcal{O}_Y$. Moreover,  in the language of \cite{Amoretti:2018tzw} we have $\omega_{gap}\sim \omega_r^2$.} $\omega_{gap}=w\,\Xi/(2\,\chi_{QQ}^2)$.

It is interesting to notice that, when $w<0$, the theory develops an instability. This can be easily seen from the square root in \eqref{eq:gap} which will become imaginary in this case. This instability is simply a mode of the system which wants to align the VEV and explicit deformation of the system in the complex plane.

Finally, we would like to flesh out some of the global effects of the explicit breaking on our observables. In order to do this, we will consider the Green's functions in equation \eqref{eq:greens_pinning} at $k=0$. In that limit,  from the Green's function $G_{\mathcal{O}_Y J^x}$ we see that the transport current $J^x$ and the operator $\mathcal{O}_Y$ completely decouple. Similarly, we will see that $J^x$ also decouples from the charge density $J^t$. It is therefore natural to expect that the small explicit breaking will have no effect on the low frequency electric conductivity,
\begin{align}
\sigma_{AC}(\omega)=\frac{G_{J^x J^x}(\omega,k=0)}{i\,\omega}=\frac{i\,\chi_{JJ}}{\omega}+\sigma_d\,.
\end{align}
However, we see that at $k=0$ the charge density and the scalar operator $\mathcal{O}_Y$ remain coupled turning the electric charge to an almost conserved quantity in the deformed theory.

\section{Numerical checks}\label{sec:numerics}

In this section we carry out a series of numerical checks of the results derived in the previous sections and in particular, we numerically confirm the sound mode dispersion relation \eqref{eq:disp_rel}, the Green's functions \eqref{eq:Greens}-\eqref{eq:mathcalG} and the formula for the gap \eqref{eq:gap}. 

We consider the action \eqref{eq:baction2} with the following potential and gauge coupling 
\begin{align} %\lambda1=c^2/2, \lambda2=-\lambda
&V=-6+m_\rho^2 \rho^2+\frac{c^2}{2}\phi^4+\frac{m_\phi^2}{2}\phi^2+ \lambda \,\rho^2\phi^2\,,\quad m_\rho^2=-2\,,\quad m^2_\phi=-2\,,\nonumber\\
&\tau=\cosh{\phi},
\end{align}
where we note that, for $\rho=0$, $\partial_\phi V=0$ both at $\phi=0$ and at $\phi=1/c$. 

Given the choices above, the corresponding equations of motion admit a unit-radius $AdS_4$ vacuum solution with $\rho =\phi= A=\theta=0$, which is dual to a $d = 3$ CFT. Placing the CFT at finite temperature corresponds to considering the Schwarzschild black hole, which typically serves as the configuration dual to the normal phase of holographic systems. However, here we choose to deform our boundary theory by a relevant operator, $O_\phi$, with scaling dimension $\Delta_\phi= 2$. Then, the corresponding backreacted solution dual to the normal phase of our system will  be given by black brane with a non-trivial profile for the scalar field $\phi$. As the temperature goes to zero, $T\to 0$, these configurations will approach a flow between the unit-radius $AdS_4$ in the UV, with $\phi=0$,  and an IR $AdS_4$ with radius $L_{IR}^2=12 c^2/(1+12 c^2)$ supported by $\phi=1/c$. 

\begin{figure}[h!]
\centering
\includegraphics[width=0.6\linewidth]{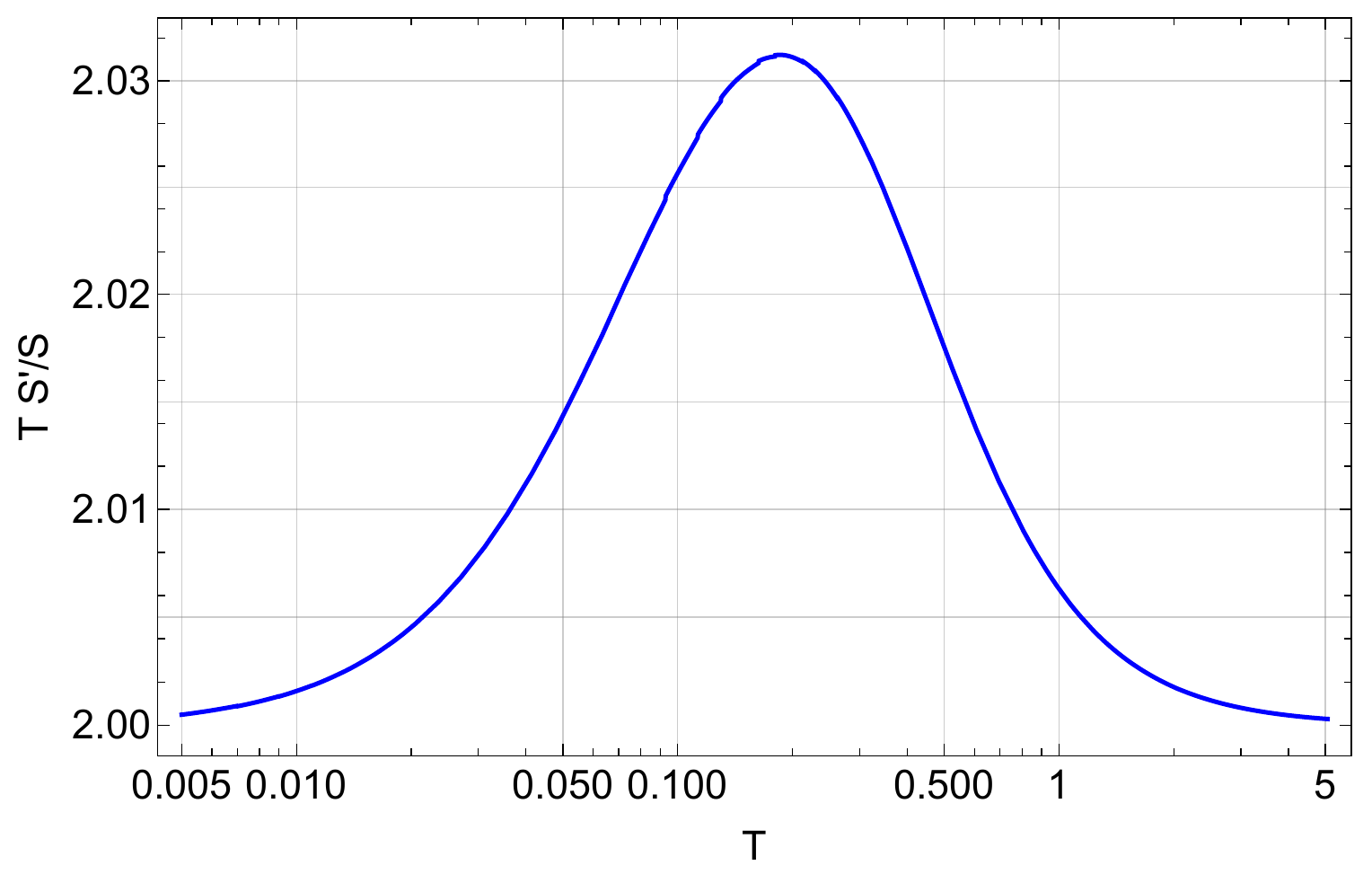}
\caption{Plot of the logarithmic derivative of the entropy $T S'(T)/S(T)$ as a function of the temperature indicating that our solutions interpolate between two $AdS_4$ geometries. Here $\phi_{(s)} = 1$ and $c = 1$.}
\label{fig:entropy}
\end{figure}

To construct these solutions explicitly we consider the ansatz \eqref{eq:background} with $\rho(r)=0$ and the IR and UV boundary conditions, \eqref{eq:bhor_exp} and \eqref{eq:UV_expansion} respectively, with $\rho^{(0)}=$$\rho_{(s)}=\rho_{(v)}=0$. This boundary condition problem is then solved using a double-sided shooting technique. In figure \ref{fig:entropy},  we plot the logarithmic derivative of the entropy of the system with respect to the temperature, $T S'(T )/S(T )$ for $\phi_{(s)}  = 1$ and $c = 1$. We clearly see that both at very high and very low temperatures the entropy scales like $T^2$ which is compatible with having $AdS_4$ on both sides of the RG flow. 

On top of these thermal states, we will consider instabilities associated to the scalar field $\rho$. To ensure that such instabilities exist in our model we need to make sure that the scalar field $\rho$ violates the BF bound associated with the $AdS_4$ in the IR, i.e.
\begin{equation}
L^2_{IR} \,{m^{IR}_\rho}^2=\frac{12}{1+12 c^2}(c^2\,m_\rho^2+\lambda)< -\frac{9}{4}\,,
\end{equation}
but is nevertheless stable in the UV: $m_\rho^2\ge-\frac{9}{4}\,$. For example, this is the case for $c=1,\lambda=-3/2$ and so we expect an instability to occur for this choice of parameters. Thus, for temperatures below a critical one, we expect a new branch of black holes to emerge characterised by a non-trivial condensate for $\rho$. To determine the critical temperature at which these instabilities set in we need to study the associated zero mode. In particular, we consider a linearised perturbation around the background constructed above of the form
\begin{align}
\rho&=0+\delta\rho\,.
\end{align}    
Plugging the above perturbation in the equations of motion, we obtain one second order linear ODE, which we solve by imposing the following boundary conditions at the black hole horizon
\begin{align} 
\delta \rho&=\delta\rho_h+\cdots\,,
\end{align}  
and asymptotically
\begin{align}
\delta \rho&=0+\frac{\delta\rho_v}{r^2}+\cdots\,,
\end{align}  
where we have already set the source for $\delta \rho$ to $0$, so that the emergence of the new phase is spontaneous. Overall, the boundary conditions are determined by 2 constants $\delta\rho_h\,,\delta\rho_v$, one of which can be set to $1$ because of the linearity of the equation. Consequently, performing the numerical integration will fix the highest value of the temperature, $T_c= 0.022$, for which one can find non-trivial solutions for $\delta\rho$. 

\begin{figure}[h!]
\centering
\includegraphics[width=0.48\linewidth]{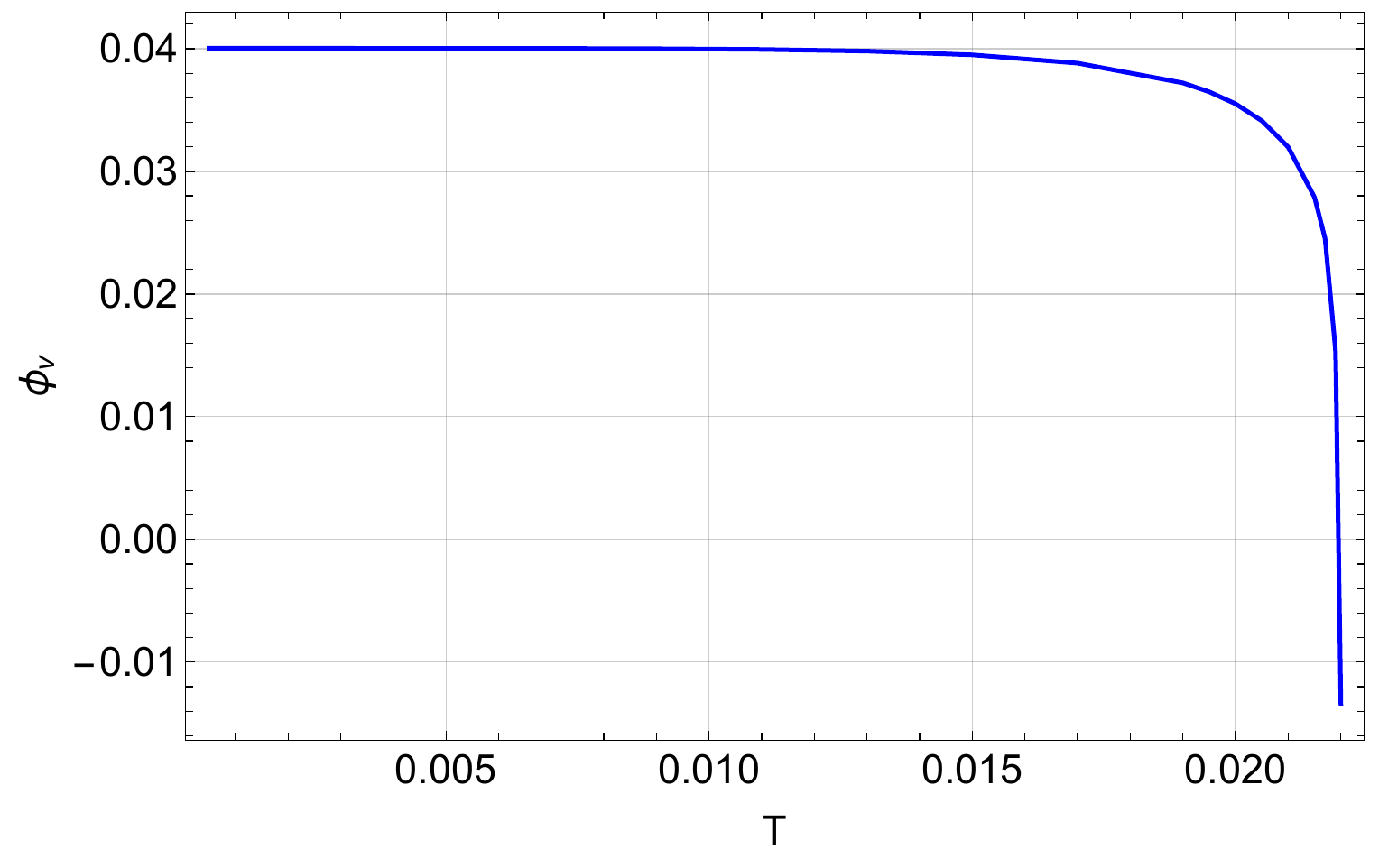}\quad\includegraphics[width=0.48\linewidth]{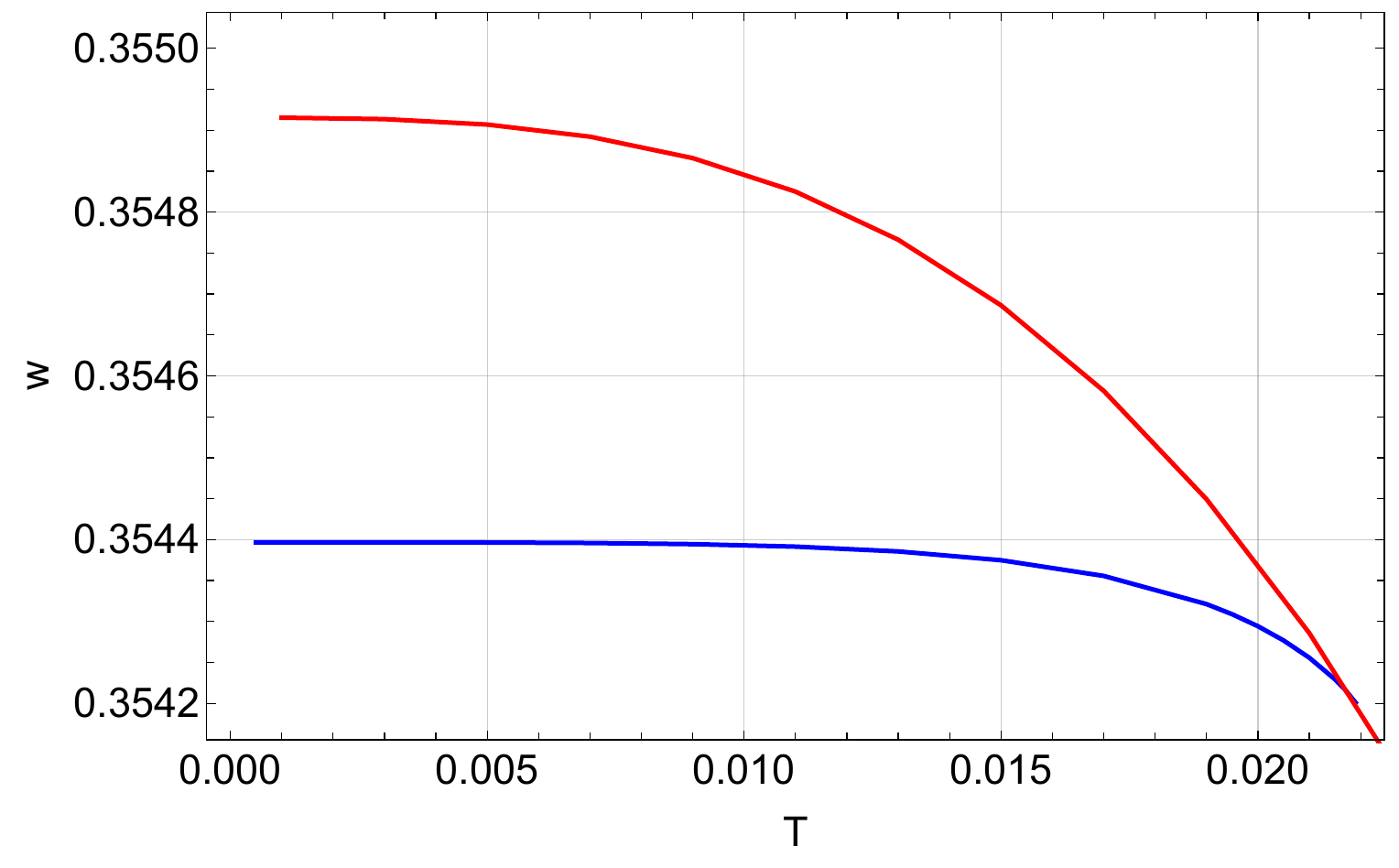}
\caption{(left) Plot of the condensate as a function of the temperature. (right) Plot of the free energy of the system in the normal phase (red line) and in the broken phase (blue line) as a function of the temperature. Both plots demonstrate that there is a second order phase transition at $T=T_c$.}
\label{fig:PT}
\end{figure}

The next step is to construct the backreacted solutions corresponding to the broken phase. To achieve this we use the ansatz \eqref{eq:background} and the IR and UV boundary conditions, \eqref{eq:bhor_exp}, \eqref{eq:UV_expansion}. When plugging this ansatz in the equations of motion we obtain a set of three second order ODEs and one first order. Thus, a solution is specified in terms of 7 constants of integration. Looking at the expansion \eqref{eq:bhor_exp}, we see that it is specified in terms of 3 constants, in addition to the temperature $T$. Similarly, the asymptotic expansion \eqref{eq:UV_expansion} is parametrised by 4 constants in addition to $\phi_{(s)}$ \textemdash note that we keep $\rho_{(s)}=0$ so that the condensation is spontaneous. Overall, in the IR and UV expansions we have a total of 7 constants as well as $T$ and $\phi_{(s)} $, which matches the 7 constants of integration. We proceed to solve this boundary condition problem numerically using double-sided shooting. In figure \ref{fig:PT} we show the condensate as well as the free energy  as functions of the temperature, $T$, in support of the phase transition being second order. Here $\phi_{(s)} =1$ and  $c=1,\lambda=-3/2$.

\subsection{Static perturbations}
Having constructed the backreacted black holes corresponding to the broken phase, we now turn our attention to studying perturbations around them. In particular, to check numerically the validity of the analytic expressions for the dispersion relation of the sound mode and the spatially resolved two-point functions, we need to construct the static perturbations as in section \ref{sec:thermo} and extract from them $\chi_{JJ}$, $\chi_{QQ}$. 

Specifically, considering the perturbations $\delta B_t^{(t)}(r)$ and $\delta B_x^{(x)}(r)$ gives rise to two \textit{decoupled} linear second order equations, which we solve subject to boundary conditions \eqref{eq:therm_pert_UV},\eqref{eq:therm_pert_IR}. %Note that for both cases we can take advantage the linearity of the equations to set $ \delta b_x = \delta b_t=1$.Then, for each of the equations, the number of integration constant matches exactly the undetermined constants and we thus have a unique solution. 
Then, given the numerical solutions, the susceptibilities are simply obtained by dividing the corresponding expectation values by the associated sources as extracted by the asymptotic expansion
\begin{equation}
\chi_{QQ}=-\delta j_t/\delta \mu_t\,,\qquad \chi_{JJ}=-\delta j_x/\delta \mu_x\,.
\end{equation}
On the other hand,  $\Xi$, $\sigma_d$ are calculated using horizon data. For $\{T,c,\lambda,\phi_{(s)}\}=\{0.015,1,-3/2,1\}$, we find  $\chi_{JJ}=0.12238$, $\chi_{QQ}=0.138$, $\Xi=0.408$, $\sigma_d=0.318$.

\subsection{Second sound and spatially resolved two-point functions}
In order to compute the second sound and the two-point functions we need to go beyond static perturbations. In particular, we consider the linearised perturbation \eqref{eq:qnm_ansatz}, with 
\begin{equation}
\label{eq:Seq}
S(r)=\int_{\infty}^{r}\frac{dy}{U(y)}\,.
\end{equation}
Plugging this ansatz in the equations of motion, we obtain two second order ODEs, along with an algebraic equation for $\delta B_r$. We now turn to the boundary conditions for these functions. In the IR we impose in-falling boundary conditions at the horizon which is located at $r=0$
\begin{align}
&\delta B_t=b_t+\dots\nonumber\\
&\delta B_{x_1}=b_{x_1}+\dots\,.\nonumber
\end{align}
Thus, we see that the expansion is fixed in terms of 2 constant $b_t, b_{x_1}$, in addition to $\omega,k$. The UV expansion takes the form
\begin{align}
\label{eq:BUV}
&\delta B_t=\delta s_t- i\omega \delta c+\frac{\delta j_t}{r+R}+\dots\nonumber\\
&\delta B_{x_1}=\delta s_{x_1}+ ik \delta c+\frac{\delta j_{x_1}}{r+R}+\dots,\,
\end{align}
where $\delta j_t$ is fixed in terms of the other parameters. Overall, this expansion is determined in terms of the sources $\delta s_t, \delta s_x$ and the 2 parameters $\delta c, \delta j_{x_1}$, in addition to $\omega, k$.

\begin{figure}[h!]
\centering
\includegraphics[width=0.48\linewidth]{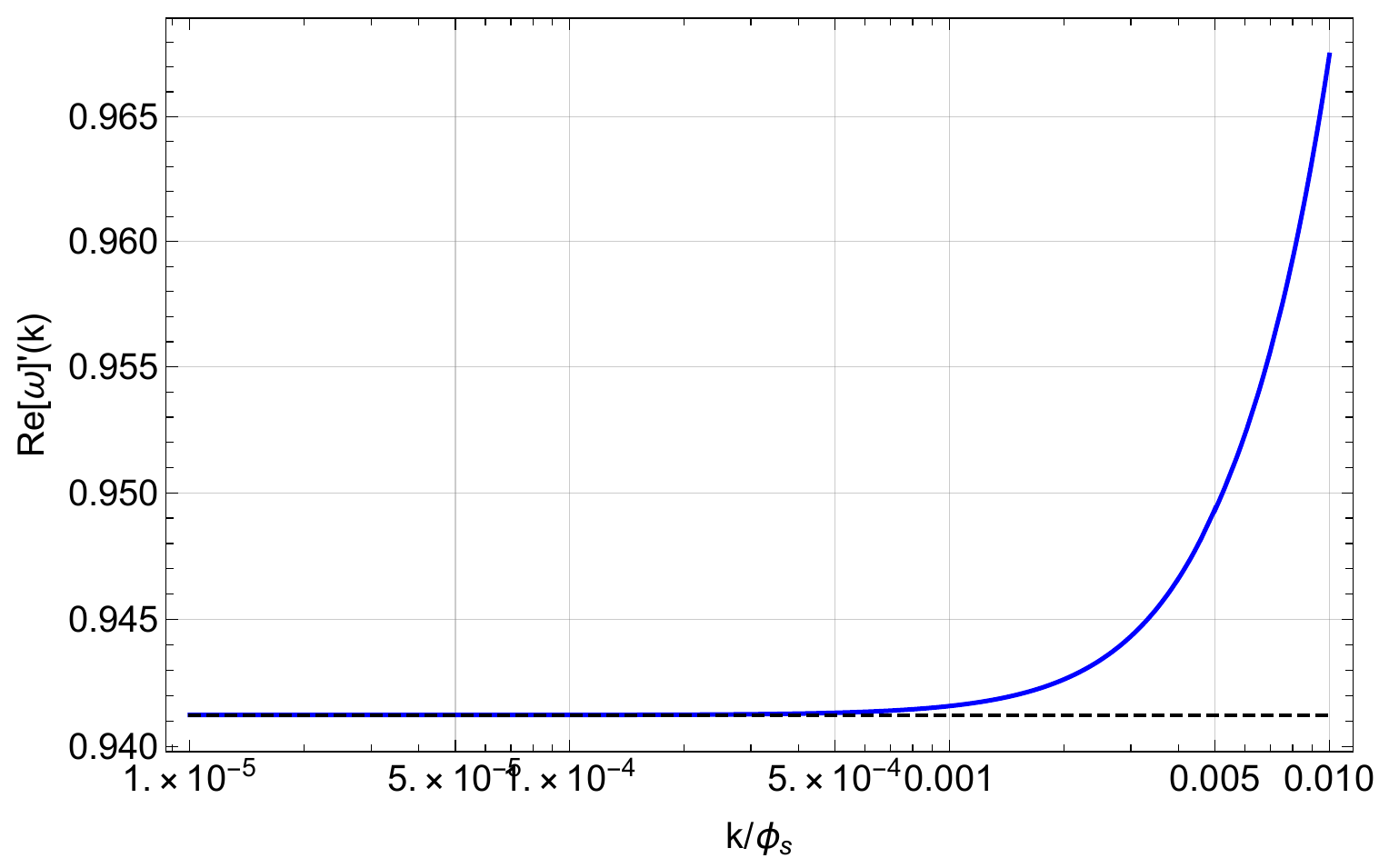}\quad\includegraphics[width=0.48\linewidth]{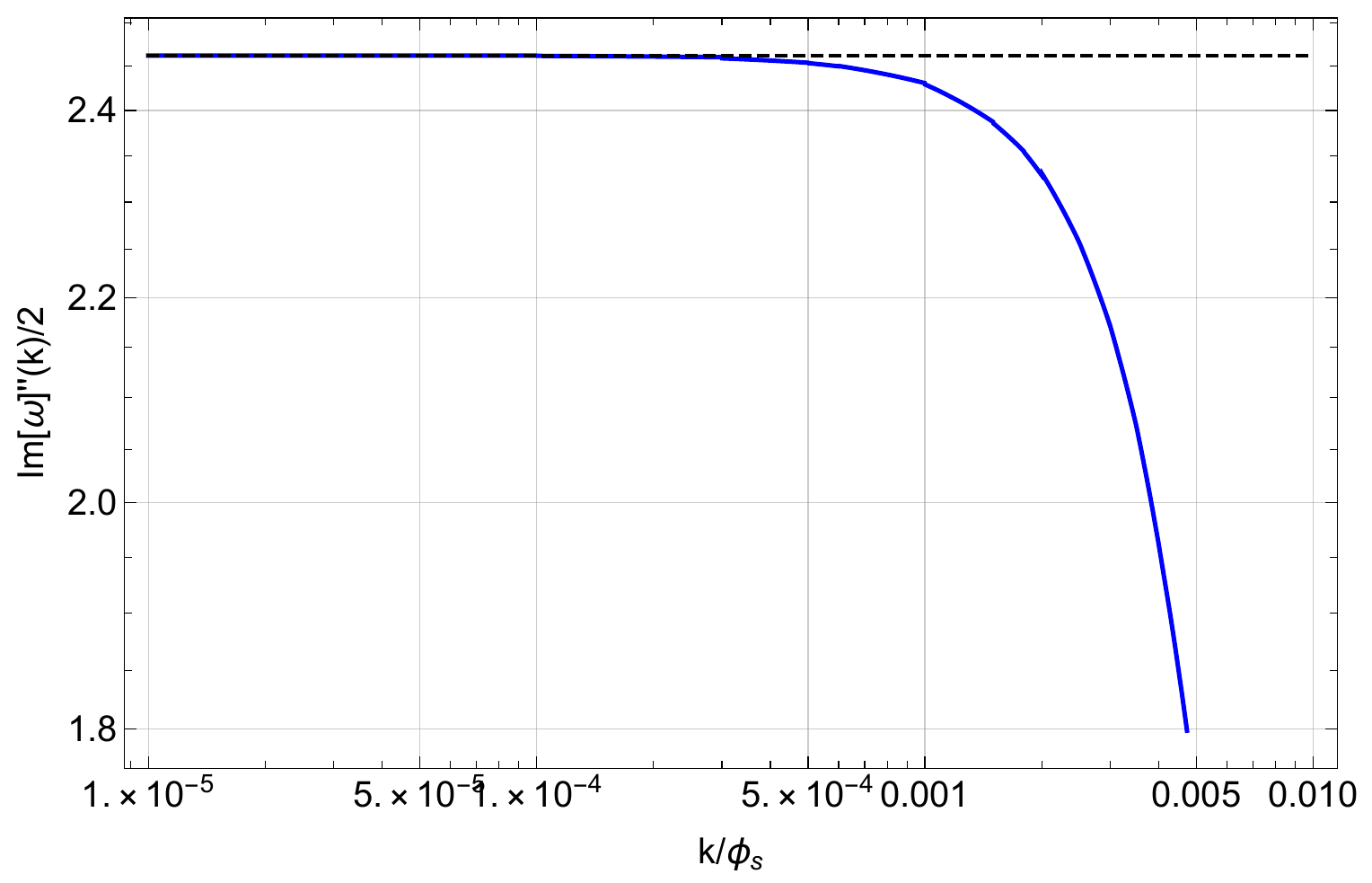}
\caption{Plots of  $\tfrac{ \omega^2 Re[\mathcal{G}]}{\partial k}$  and $\frac{1}{2}\frac{\partial^2 Im[\omega]}{\partial k^2}$  as functions of $k$ for the second sound dispersion relation. The dashed black line corresponds to the analytic prediction. Here $\{T,c,\lambda,\phi_{(s)}\}=\{0.015,1,-3/2,1\}$. }
\label{fig:SS}
\end{figure}

For constructing the second sound, we solve the above equations around the numerical background of the previous section, imposing that the sources vanish $\delta s_t=\delta s_{x_1}=0$. In addition, due to the linearity of the equations we also impose $b_{x_1}=1$. Thus, for fixed $k$, the shooting method determines $b_t, \delta j_{x_1},\delta c, \omega$. In figure \ref{fig:SS}, we compare our numerical results with the analytics of the previous chapter by plotting certain derivatives of the dispersion relation. For small values of $k$, where our analytic arguments are valid, we see a good quantitative agreement.

For constructing the two point functions we again employ a double-sided shooting technique, but now we set either $\delta s_t$ or $\delta s_{x_1}$ to zero  and scale the remaining source to one using the linearity of the equations. Thus, for fixed $\omega, k$ and say $\delta s_t=0, \delta s_x=1$, the boundary condition system determines $ b_t, b_{x_1}, \delta c, \delta j_x$, allowing one to compute $G_{xx}= \omega^2 \,\mathcal{G}(k,\omega), G_{tx}=G_{xt}=k \omega \,\mathcal{G}(k,\omega), G_{tt}= k^2 \,\mathcal{G}(k,\omega)$ where

\begin{equation}
\mathcal{G}(k,\omega)=\frac{ \delta j_x+i \omega- k \omega \delta c}{\omega^2}\,.
\end{equation}

\begin{figure}[h!]
\centering
\includegraphics[width=0.48\linewidth]{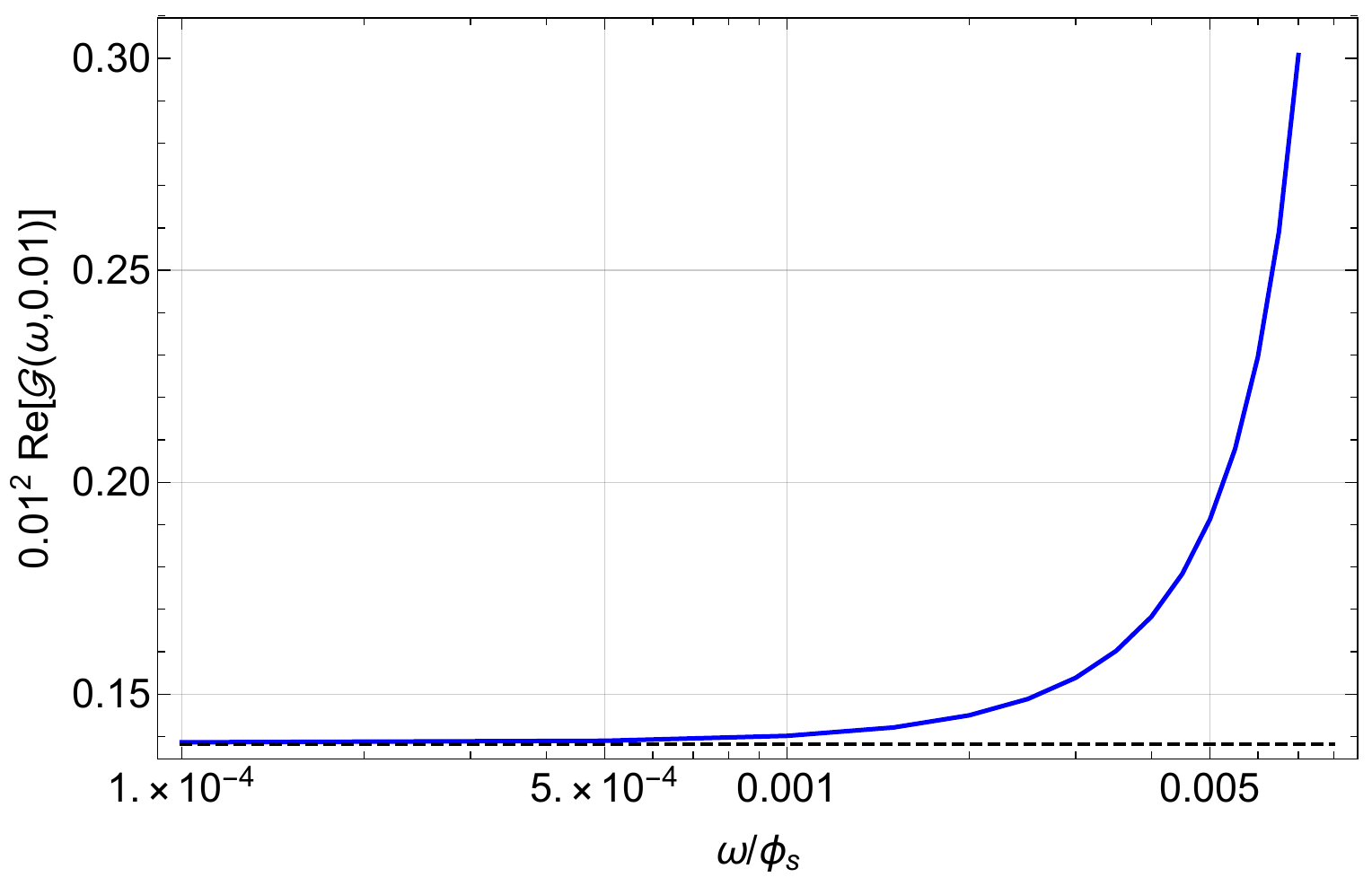}\quad\includegraphics[width=0.48\linewidth]{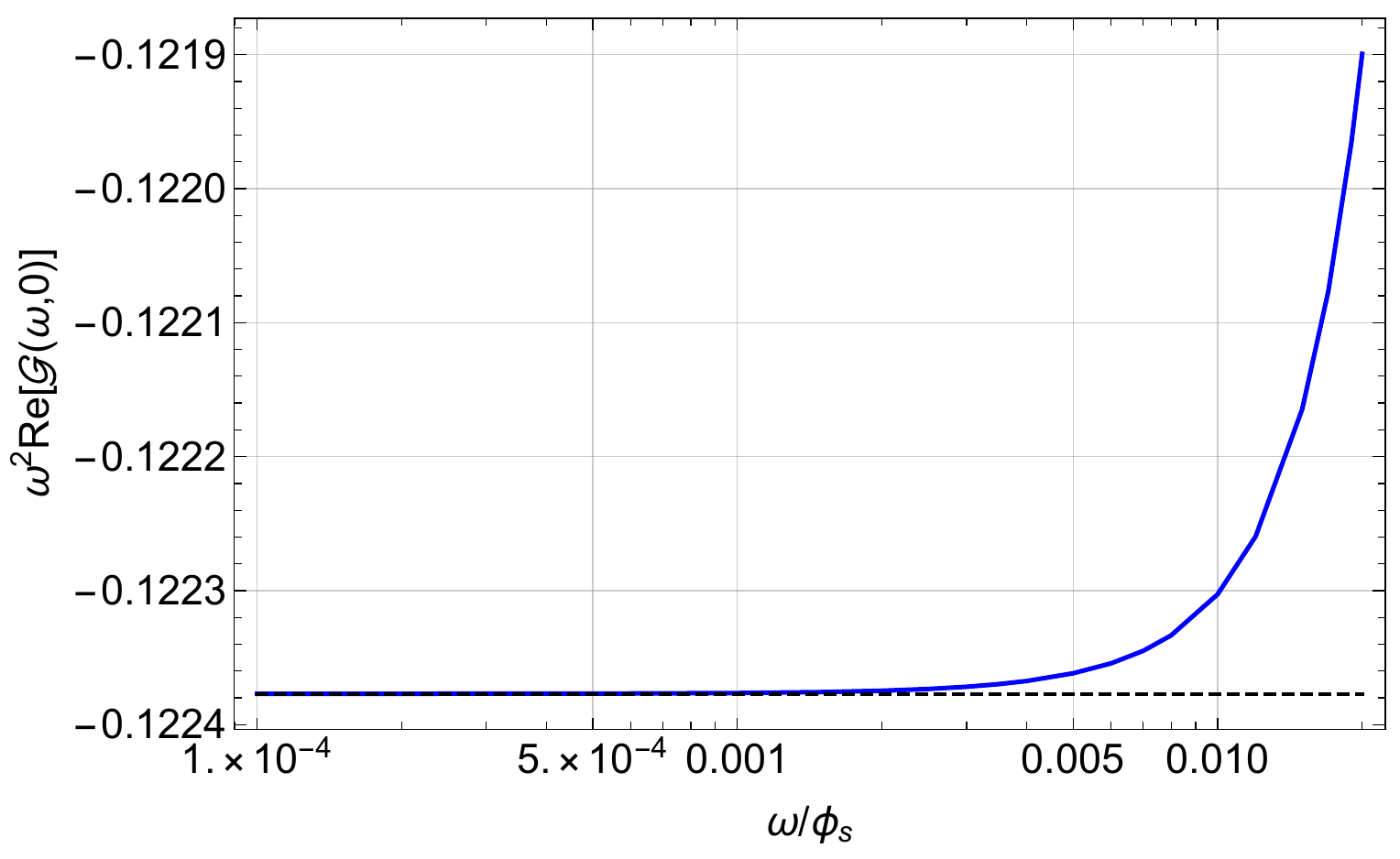}\\
\includegraphics[width=0.48\linewidth]{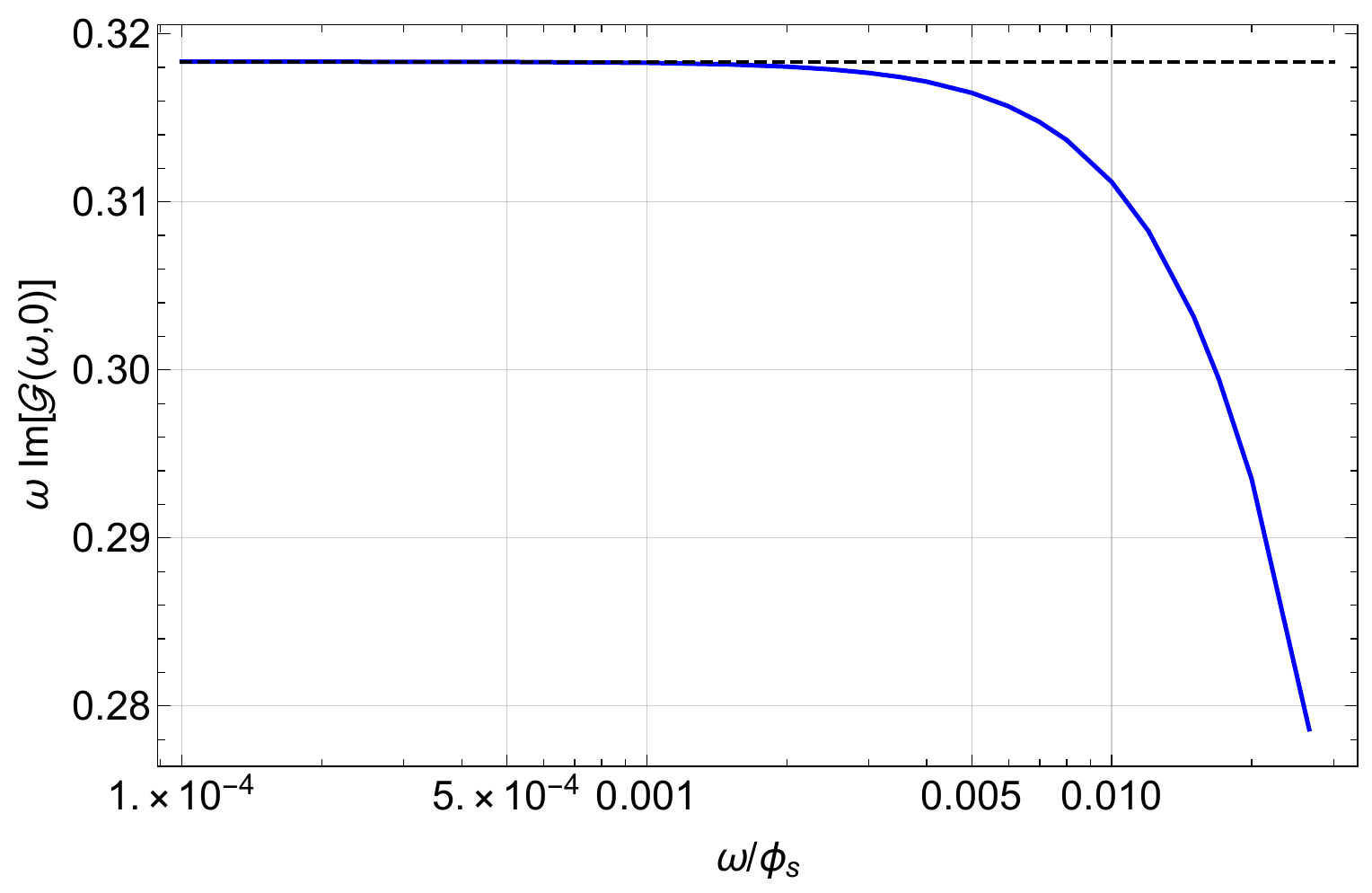}\quad\includegraphics[width=0.48\linewidth]{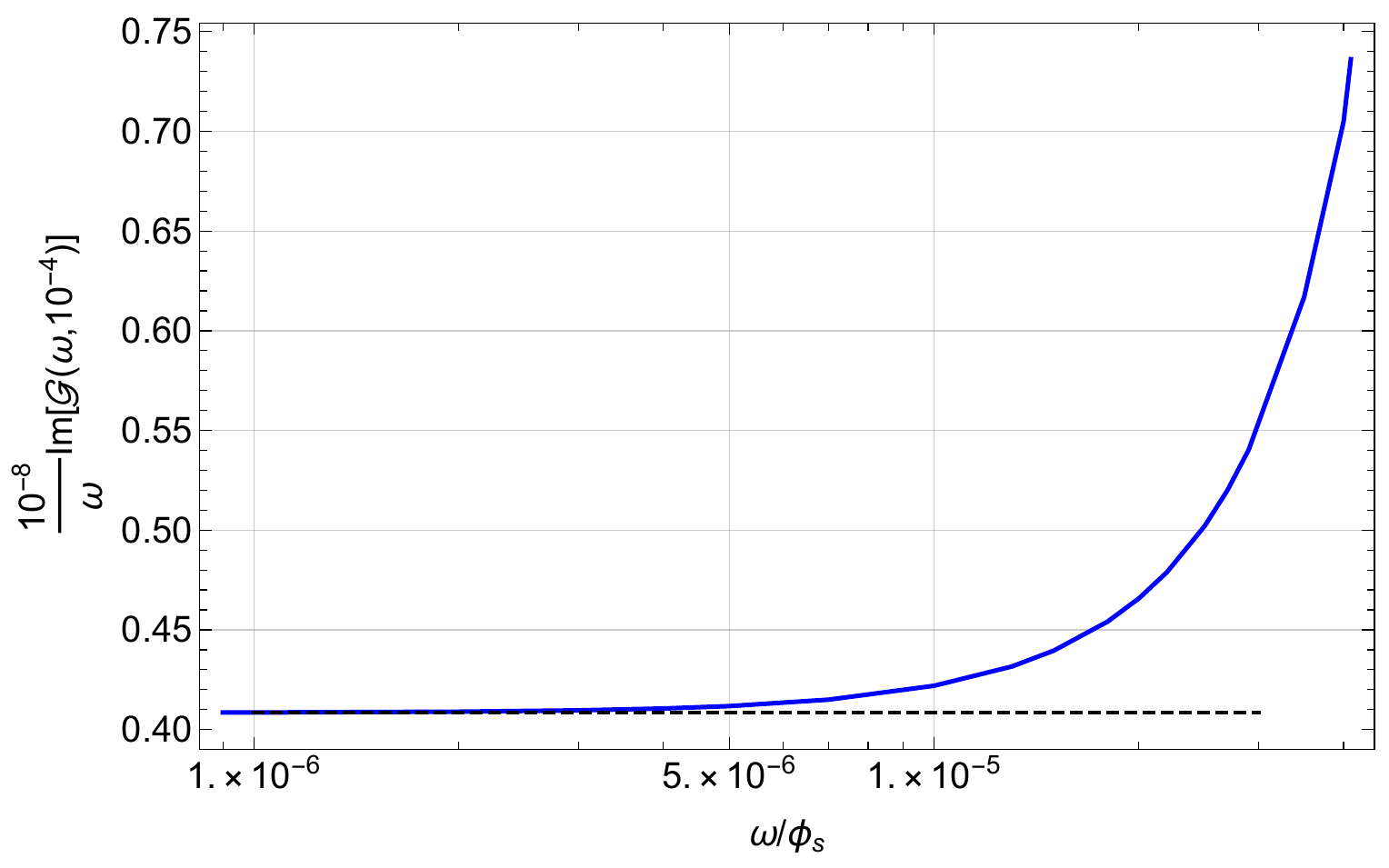}
\caption{Plots of $\mathcal{G}$  as a function of $\omega$ for $k=0$ and $k=$ finite. The dashed black lines corresponds to  $\chi_{QQ}$ (top, left),  $-\chi_{JJ}$ (top, right), $\sigma_d$ (bottom, left) and $\Xi$ (bottom, right). Here $\{T,c,\lambda,\phi_{(s)}\}=\{0.015,1,-3/2,1\}$.}
\label{fig:traspCoeff}
\end{figure}

In figure \ref{fig:traspCoeff} we plot $\mathcal{G}$ (scaled appropriately) as function of the frequency, either for zero and finite $k$. In the small frequency limit, these quantities approach the constants $\chi_{QQ}$ (top, left),  $-\chi_{JJ}$ (top, right), $\sigma_d$ (bottom, left) and $\Xi$ (bottom, right), in agreement with the analytic expressions  \eqref{eq:Greens}-\eqref{eq:mathcalG}.

\subsection{Pseudo-gapless modes}
\begin{figure}[h!]
\centering
\includegraphics[width=0.48\linewidth]{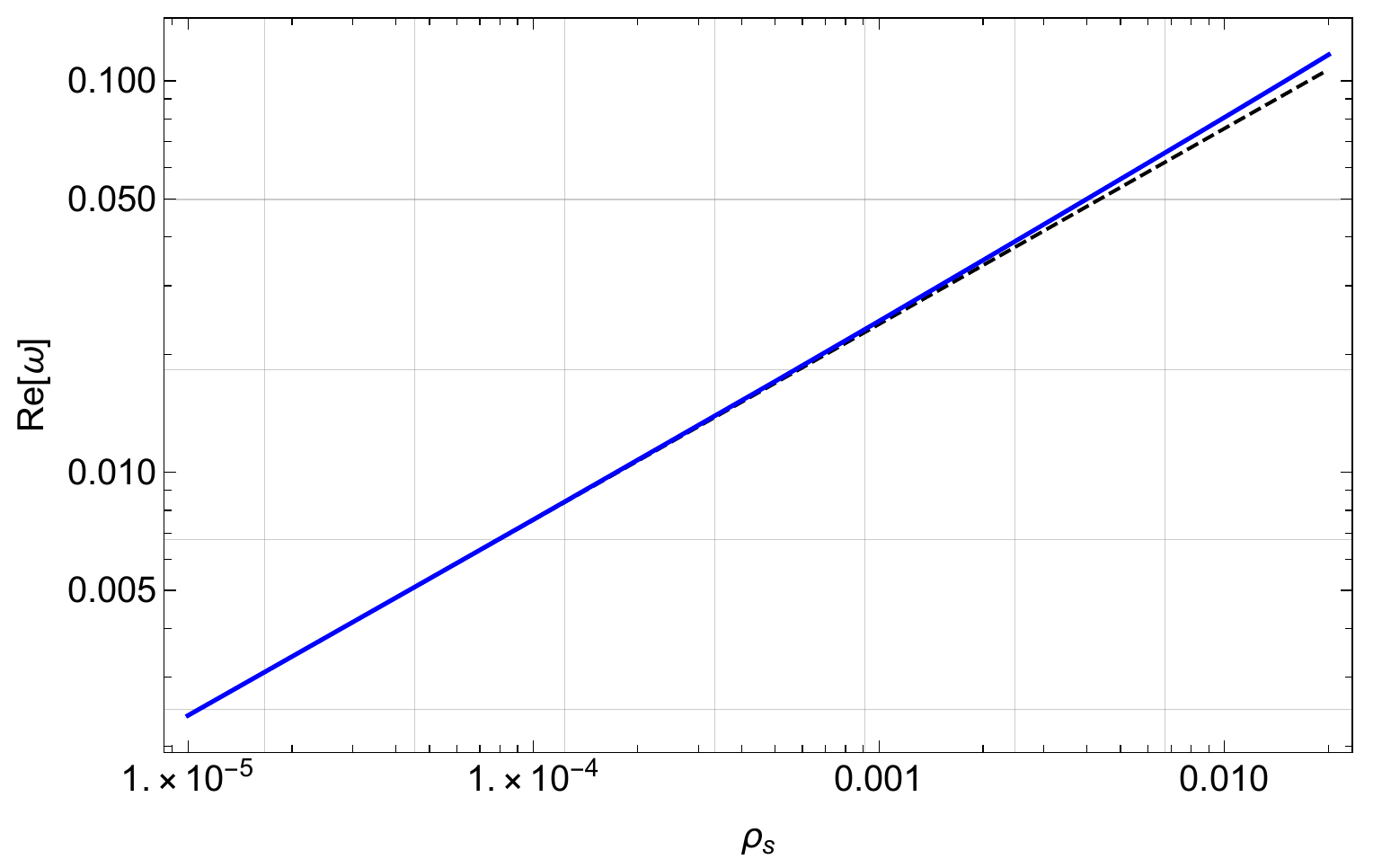}\quad\includegraphics[width=0.48\linewidth]{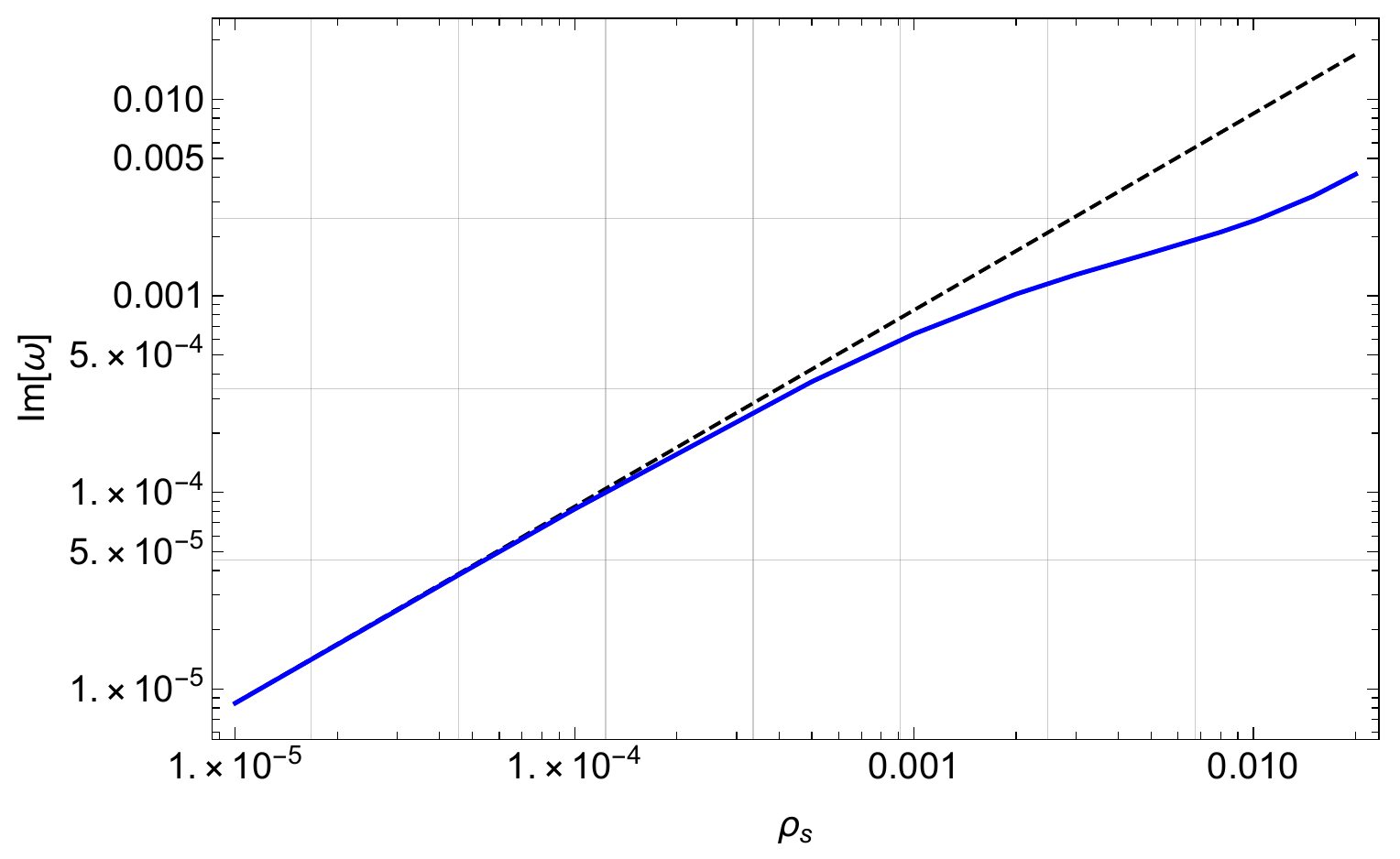}\\
\caption{Plots of the real and imaginary part of the gap as function of the source $\rho_{(s)}$. The dashed black lines corresponds to the analytic prediction. Here $\{T, c,\lambda,\phi_{(s)}\}=\{0.015,1,-3/2,1\}$ and $k=10^{-4}$.}
\label{fig:gap}
\end{figure}

In this subsection we outline the numerical computation of the pseudo-gapless modes in the presence of
pinning. We perform a calculation similar to the one for second sound, but we now consider linearised fluctuations \eqref{eq:qnm_ansatz}, \eqref{eq:Seq} around a background configuration that has a small but finite source, $\rho_{(s)}$, for the scalar $\rho$. The only difference to the previous subsection is the expansion of the perturbations in the UV part of the geometry. In this case we find that
\begin{align}
&\delta B_t=0+\frac{\delta j_t}{r+R}+\dots\nonumber\\
&\delta B_{x_1}=0+\frac{\delta j_{x_1}}{r+R}+\dots\,.
\end{align}
Note that this expansion differs from \eqref{eq:BUV}, not only because there are no sources for the perturbations in this case, but also because having $\rho_{(s)}\ne0$ pushes the VEV of the goldstone mode to appear at order $1/(r+R)$ leading to  $\delta j_t,\delta j_{x_1}$ being independent constants. The parameter counting follows just like above, suggesting that for fixed $k$ we expect to find a discrete set of solutions. In figure \ref{fig:gap}, we plot our numerical results for the real and imaginary part of the gap and we overlay them with the analytics of the previous chapter. We see a good quantitative agreements for small values of  $\rho_{(s)}$, as expected.

\section{Discussion}

In this paper we studied first order dissipative effects in the hydrodynamic regime of holographic superfluid phases of matter. At zero chemical potential and charge density, the normal and the superfluid collective degrees of freedom remain decoupled. This fact simplified our analysis in extracting the transport coefficients relevant to the superfluid dissipation.

For a relativistic superfluid, in principle we would only have to determine four invariant quantities that would fully fix the constitutive relation of the conserved current in terms of the phase \cite{KHALATNIKOV198270}. In our case, only two of them were non-trivial and they were both fixed in terms of susceptibilities and black hole horizon data. For a specific class of models the coefficient $\sigma_d$ had been computed earlier in the literature \cite{Davison:2015taa,Gouteraux:2019kuy}. Here, we presented a technique which is applicable also in inhomogeneous black hole backgrounds relevant to phases of matter in which translations are broken either spontaneously or explicitly. Interesting extensions of our work include the reduced hydrodynamics of superfluid phases without momentum conservation \cite{toapp}.

One interesting aspect of our work concerns the coefficient $\Xi$ which appears in the time component of constitutive relations \eqref{eq:jconst_rel_sources} for the current. We have shown that this grows like $(T_c-T)^{-1}$ close to the transition signalling the breakdown of the derivative expansion.  More specifically, judging from the next to leading term in the hydrodynamic expansion, we have made explicit that this will converge for wavenumbers with $k\ll T_c-T$.  It would be interesting to explore the precise way that the hydrodynanic expansion breaks down by following the logic of e.g. \cite{Grozdanov:2019uhi,Grozdanov:2019kge,Withers:2018srf}. Finally, we have studied the hydrodynamics of the system upon introducing sources which break the global symmetry in a controlled manner.

An interesting extension we will report on in the future is the inclusion of a background finite magnetic field \cite{toapp}. That would imply the existence of background vortices which have been previously studied in the framework of holography in \cite{Donos:2020viz}. One step further would include the presence of disorder and the resulting flux pinning which relaxes the supercurrent leading to finite DC electric conductivity at finite temperature.

Finally, holography provides access to a plethora of superfluid phase ground states \cite{Gubser:2009gp,Gouteraux:2012yr}.  An interesting direction would be to study the low temperature behaviour of the transport coefficients we computed in this paper. This is possible by extracting the behaviour of low temperature black hole horizons by following the techniques of e.g. \cite{Donos:2014uba}.

\section*{Acknowledgements}

AD is supported by STFC grant ST/T000708/1. CP is supported by the European Union’s Horizon 2020 research and innovation programme under the Marie Skłodowska-Curie grant agreement HoloLif No 838644. 

\newpage
\bibliographystyle{utphys}
\bibliography{refs}{}

\providecommand{\href}[2]{#2}\begingroup\raggedright\begin{thebibliography}{10}

\bibitem{Gubser:2008px}
S.~S. Gubser, ``{Breaking an Abelian gauge symmetry near a black hole
  horizon},'' \href{http://dx.doi.org/10.1103/PhysRevD.78.065034}{{\em Phys.
  Rev.} {\bfseries D78} (2008) 065034},
\href{http://arxiv.org/abs/0801.2977}{{\ttfamily arXiv:0801.2977 [hep-th]}}.
%%CITATION = ARXIV:0801.2977;%%.

\bibitem{Hartnoll:2008kx}
S.~A. Hartnoll, C.~P. Herzog, and G.~T. Horowitz, ``{Holographic
  Superconductors},''
  \href{http://dx.doi.org/10.1088/1126-6708/2008/12/015}{{\em JHEP} {\bfseries
  12} (2008) 015},
\href{http://arxiv.org/abs/0810.1563}{{\ttfamily arXiv:0810.1563 [hep-th]}}.
%%CITATION = ARXIV:0810.1563;%%.

\bibitem{Hartnoll:2008vx}
S.~A. Hartnoll, C.~P. Herzog, and G.~T. Horowitz, ``{Building a Holographic
  Superconductor},''
  \href{http://dx.doi.org/10.1103/PhysRevLett.101.031601}{{\em Phys. Rev.
  Lett.} {\bfseries 101} (2008) 031601},
\href{http://arxiv.org/abs/0803.3295}{{\ttfamily arXiv:0803.3295 [hep-th]}}.
%%CITATION = ARXIV:0803.3295;%%.

\bibitem{PhysRev.60.356}
L.~Landau, ``Theory of the superfluidity of helium ii,''
  \href{http://dx.doi.org/10.1103/PhysRev.60.356}{{\em Phys. Rev.} {\bfseries
  60} (Aug, 1941) 356--358}.
  \url{https://link.aps.org/doi/10.1103/PhysRev.60.356}.

\bibitem{PhysRev.72.838}
L.~Tisza, ``The theory of liquid helium,''
  \href{http://dx.doi.org/10.1103/PhysRev.72.838}{{\em Phys. Rev.} {\bfseries
  72} (Nov, 1947) 838--854}.
  \url{https://link.aps.org/doi/10.1103/PhysRev.72.838}.

\bibitem{KHALATNIKOV198270}
I.~Khalatnikov and V.~Lebedev, ``Relativistic hydrodynamics of a superfluid
  liquid,''
  \href{http://dx.doi.org/https://doi.org/10.1016/0375-9601(82)90268-7}{{\em
  Physics Letters A} {\bfseries 91} no.~2, (1982) 70--72}.
  \url{https://www.sciencedirect.com/science/article/pii/0375960182902687}.

\bibitem{ISRAEL198179}
W.~Israel, ``Covariant superfluid mechanics,''
  \href{http://dx.doi.org/https://doi.org/10.1016/0375-9601(81)90169-9}{{\em
  Physics Letters A} {\bfseries 86} no.~2, (1981) 79--81}.
  \url{https://www.sciencedirect.com/science/article/pii/0375960181901699}.

\bibitem{Bhattacharya:2011eea}
J.~Bhattacharya, S.~Bhattacharyya, and S.~Minwalla, ``{Dissipative Superfluid
  dynamics from gravity},''
  \href{http://dx.doi.org/10.1007/JHEP04(2011)125}{{\em JHEP} {\bfseries 1104}
  (2011) 125},
\href{http://arxiv.org/abs/1101.3332}{{\ttfamily arXiv:1101.3332 [hep-th]}}.
%%CITATION = ARXIV:1101.3332;%%.

\bibitem{Davison:2016hno}
R.~A. Davison, L.~V. Delacr{\'e}taz, B.~Gout{\'e}raux, and S.~A. Hartnoll,
  ``{Hydrodynamic theory of quantum fluctuating superconductivity},''
  \href{http://dx.doi.org/10.1103/PhysRevB.96.059902,
  10.1103/PhysRevB.94.054502}{{\em Phys. Rev.} {\bfseries B94} no.~5, (2016)
  054502}, \href{http://arxiv.org/abs/1602.08171}{{\ttfamily arXiv:1602.08171
  [cond-mat.supr-con]}}.
[Erratum: Phys. Rev.B96,no.5,059902(2017)].
%%CITATION = ARXIV:1602.08171;%%.

\bibitem{Crnkovic:1986ex}
C.~Crnkovic and E.~Witten, ``{Covariant description of canonical formalism in
  geometrical theories},''.

\bibitem{Andrade:2015iyf}
T.~Andrade and A.~Krikun, ``{Commensurability effects in holographic
  homogeneous lattices},''
  \href{http://dx.doi.org/10.1007/JHEP05(2016)039}{{\em JHEP} {\bfseries 05}
  (2016) 039}, \href{http://arxiv.org/abs/1512.02465}{{\ttfamily
  arXiv:1512.02465 [hep-th]}}.

\bibitem{Amoretti:2018tzw}
A.~Amoretti, D.~Are{\'a}n, B.~Gout{\'e}raux, and D.~Musso, ``{Universal
  relaxation in a holographic metallic density wave phase},''
  \href{http://dx.doi.org/10.1103/PhysRevLett.123.211602}{{\em Phys. Rev.
  Lett.} {\bfseries 123} no.~21, (2019) 211602},
\href{http://arxiv.org/abs/1812.08118}{{\ttfamily arXiv:1812.08118 [hep-th]}}.
%%CITATION = ARXIV:1812.08118;%%.

\bibitem{Donos:2019hpp}
A.~Donos, D.~Martin, C.~Pantelidou, and V.~Ziogas, ``{Incoherent hydrodynamics
  and density waves},'' \href{http://dx.doi.org/10.1088/1361-6382/ab6036}{{\em
  Class. Quant. Grav.} {\bfseries 37} no.~4, (2020) 045005},
  \href{http://arxiv.org/abs/1906.03132}{{\ttfamily arXiv:1906.03132
  [hep-th]}}.

\bibitem{Donos:2019txg}
A.~Donos, D.~Martin, C.~Pantelidou, and V.~Ziogas, ``{Hydrodynamics of broken
  global symmetries in the bulk},''
  \href{http://dx.doi.org/10.1007/JHEP10(2019)218}{{\em JHEP} {\bfseries 10}
  (2019) 218},
\href{http://arxiv.org/abs/1905.00398}{{\ttfamily arXiv:1905.00398 [hep-th]}}.
%%CITATION = ARXIV:1905.00398;%%.

\bibitem{Kovtun:2003wp}
P.~Kovtun, D.~T. Son, and A.~O. Starinets, ``{Holography and hydrodynamics:
  Diffusion on stretched horizons},''
  \href{http://dx.doi.org/10.1088/1126-6708/2003/10/064}{{\em JHEP} {\bfseries
  10} (2003) 064},
\href{http://arxiv.org/abs/hep-th/0309213}{{\ttfamily arXiv:hep-th/0309213
  [hep-th]}}.
%%CITATION = HEP-TH/0309213;%%.

\bibitem{Saremi:2007dn}
O.~Saremi, ``{Shear waves, sound waves on a shimmering horizon},''
  \href{http://arxiv.org/abs/hep-th/0703170}{{\ttfamily arXiv:hep-th/0703170}}.

\bibitem{Iqbal:2008by}
N.~Iqbal and H.~Liu, ``{Universality of the hydrodynamic limit in AdS/CFT and
  the membrane paradigm},''
  \href{http://dx.doi.org/10.1103/PhysRevD.79.025023}{{\em Phys.Rev.}
  {\bfseries D79} (2009) 025023},
\href{http://arxiv.org/abs/0809.3808}{{\ttfamily arXiv:0809.3808 [hep-th]}}.
%%CITATION = ARXIV:0809.3808;%%.

\bibitem{Damle:1997rxu}
K.~Damle and S.~Sachdev, ``{Nonzero-temperature transport near quantum critical
  points},'' \href{http://dx.doi.org/10.1103/PhysRevB.56.8714}{{\em Phys. Rev.}
  {\bfseries B56} no.~14, (1997) 8714},
\href{http://arxiv.org/abs/cond-mat/9705206}{{\ttfamily arXiv:cond-mat/9705206
  [cond-mat.str-el]}}.
%%CITATION = COND-MAT/9705206;%%.

\bibitem{Hartnoll:2007ih}
S.~A. Hartnoll, P.~K. Kovtun, M.~Muller, and S.~Sachdev, ``{Theory of the
  Nernst effect near quantum phase transitions in condensed matter, and in
  dyonic black holes},''
  \href{http://dx.doi.org/10.1103/PhysRevB.76.144502}{{\em Phys. Rev.}
  {\bfseries B76} (2007) 144502},
\href{http://arxiv.org/abs/0706.3215}{{\ttfamily arXiv:0706.3215
  [cond-mat.str-el]}}.
%%CITATION = 0706.3215;%%.

\bibitem{Davison:2015taa}
R.~A. Davison, B.~Gout{\'e}raux, and S.~A. Hartnoll, ``{Incoherent transport in
  clean quantum critical metals},''
  \href{http://dx.doi.org/10.1007/JHEP10(2015)112}{{\em JHEP} {\bfseries 10}
  (2015) 112},
\href{http://arxiv.org/abs/1507.07137}{{\ttfamily arXiv:1507.07137 [hep-th]}}.
%%CITATION = ARXIV:1507.07137;%%.

\bibitem{Bhaseen:2012gg}
M.~J. Bhaseen, J.~P. Gauntlett, B.~D. Simons, J.~Sonner, and T.~Wiseman,
  ``{Holographic Superfluids and the Dynamics of Symmetry Breaking},''
  \href{http://dx.doi.org/10.1103/PhysRevLett.110.015301}{{\em Phys. Rev.
  Lett.} {\bfseries 110} no.~1, (2013) 015301},
  \href{http://arxiv.org/abs/1207.4194}{{\ttfamily arXiv:1207.4194 [hep-th]}}.

\bibitem{Gouteraux:2019kuy}
B.~Gout\'eraux and E.~Mefford, ``{Normal charge densities in quantum critical
  superfluids},'' \href{http://dx.doi.org/10.1103/PhysRevLett.124.161604}{{\em
  Phys. Rev. Lett.} {\bfseries 124} no.~16, (2020) 161604},
  \href{http://arxiv.org/abs/1912.08849}{{\ttfamily arXiv:1912.08849
  [hep-th]}}.

\bibitem{toapp}
A.~Donos, P.~Kailidis, and C.~Pantelidou, ``Work in progress,''.

\bibitem{Grozdanov:2019uhi}
S.~Grozdanov, P.~K. Kovtun, A.~O. Starinets, and P.~Tadi\'c, ``{The complex
  life of hydrodynamic modes},''
  \href{http://dx.doi.org/10.1007/JHEP11(2019)097}{{\em JHEP} {\bfseries 11}
  (2019) 097}, \href{http://arxiv.org/abs/1904.12862}{{\ttfamily
  arXiv:1904.12862 [hep-th]}}.

\bibitem{Grozdanov:2019kge}
S.~Grozdanov, P.~K. Kovtun, A.~O. Starinets, and P.~Tadi{\'c}, ``{Convergence
  of the Gradient Expansion in Hydrodynamics},''
  \href{http://dx.doi.org/10.1103/PhysRevLett.122.251601}{{\em Phys. Rev.
  Lett.} {\bfseries 122} no.~25, (2019) 251601},
\href{http://arxiv.org/abs/1904.01018}{{\ttfamily arXiv:1904.01018 [hep-th]}}.
%%CITATION = ARXIV:1904.01018;%%.

\bibitem{Withers:2018srf}
B.~Withers, ``{Short-lived modes from hydrodynamic dispersion relations},''
  \href{http://dx.doi.org/10.1007/JHEP06(2018)059}{{\em JHEP} {\bfseries 06}
  (2018) 059}, \href{http://arxiv.org/abs/1803.08058}{{\ttfamily
  arXiv:1803.08058 [hep-th]}}.

\bibitem{Donos:2020viz}
A.~Donos, J.~P. Gauntlett, and C.~Pantelidou, ``{Holographic Abrikosov
  Lattices},'' \href{http://dx.doi.org/10.1007/JHEP07(2020)095}{{\em JHEP}
  {\bfseries 07} (2020) 095}, \href{http://arxiv.org/abs/2001.11510}{{\ttfamily
  arXiv:2001.11510 [hep-th]}}.

\bibitem{Gubser:2009gp}
S.~S. Gubser, S.~S. Pufu, and F.~D. Rocha, ``{Quantum critical superconductors
  in string theory and M-theory},''
  \href{http://dx.doi.org/10.1016/j.physletb.2009.12.017}{{\em Phys. Lett.}
  {\bfseries B683} (2010) 201--204},
\href{http://arxiv.org/abs/0908.0011}{{\ttfamily arXiv:0908.0011 [hep-th]}}.
%%CITATION = ARXIV:0908.0011;%%.

\bibitem{Gouteraux:2012yr}
B.~Gouteraux and E.~Kiritsis, ``{Quantum critical lines in holographic phases
  with (un)broken symmetry},''
  \href{http://dx.doi.org/10.1007/JHEP04(2013)053}{{\em JHEP} {\bfseries 04}
  (2013) 053}, \href{http://arxiv.org/abs/1212.2625}{{\ttfamily arXiv:1212.2625
  [hep-th]}}.

\bibitem{Donos:2014uba}
A.~Donos and J.~P. Gauntlett, ``{Novel metals and insulators from
  holography},'' \href{http://dx.doi.org/10.1007/JHEP06(2014)007}{{\em JHEP}
  {\bfseries 06} (2014) 007},
\href{http://arxiv.org/abs/1401.5077}{{\ttfamily arXiv:1401.5077 [hep-th]}}.
%%CITATION = ARXIV:1401.5077;%%.

\end{thebibliography}\endgroup
\end{document}